\documentclass[journal,a4paper]{IEEEtran}

\IEEEoverridecommandlockouts
\usepackage{cite}
\usepackage{amsmath,amssymb,amsfonts}
\usepackage{graphicx}
\usepackage{textcomp}
\usepackage{xcolor}
\usepackage{algorithm}
\usepackage{algorithmic}
\usepackage{amsmath}
\usepackage{graphics}
\usepackage{epsfig}
\usepackage{float}
\usepackage{bm}
\usepackage[section]{placeins}

  \newcommand{\algorithmicbreak}{\textbf{break}}
  \newcommand{\BREAK}{\STATE \algorithmicbreak}
\def\BibTeX{{\rm B\kern-.05em{\sc i\kern-.025em b}\kern-.08em
    T\kern-.1667em\lower.7ex\hbox{E}\kern-.125emX}}

\begin{document}

\title{Manifold Optimization Based Multi-user Rate Maximization Aided by Intelligent Reflecting Surface}

\author{Liyue~Zhang,~\IEEEmembership{Student Member,~IEEE,}
        Qing~Wang,~\IEEEmembership{Member,~IEEE,}
        Haozhi~Wang,
        Peng~Chen,
        Hua~Chen,\\
        Wei~Liu,~\IEEEmembership{Senior Member,~IEEE},
        Zhiqiang~Wu,~\IEEEmembership{Senior Member,~IEEE}

\thanks{This work is supported by the National Natural Science Foundation of China under Grant 61871282, Grant 62001256 and Grant U20A20162. Part of this paper has been published in the 10th IEEE/CIC International Conference on Communications in China (ICCC2021)\cite{9580434}.  (\textit{Corresponding author: Peng Chen.)}}
%\thanks{L.Y. Zhang, Q. Wang, H.Z. Wang are with The School of Electrical and Information Engineering, Tianjin University, Tianjin,
%300072 China (email: zhangliyue@tju.edu.cn; wangq@tju.edu.cn; wanghaozhi@tju.edu.cn).}
\thanks{L.Y. Zhang, Q. Wang, H.Z. Wang, P. Chen are with The School of Electrical and Information Engineering, Tianjin University, Tianjin,
300072 China (email: zhangliyue@tju.edu.cn; wangq@tju.edu.cn; wanghaozhi@tju.edu.cn; chenpeng\_1997@tju.edu.cn).}
\thanks{H. Chen is The Faculty of Information Science and Engineering, Ningbo University, Ningbo 315211, China (email: dkchenhua0714@hotmail.com).}
\thanks{W. Liu is with The Department of Electronic and Electrical Engineering, The University of Sheffield, Sheffield S1 3JD, UK (email: w.liu@sheffield.ac.uk).}
\thanks{Zhiqiang Wu is with the Department of Electrical Engineering, Tibet
University, Lhasa 850000, China (e-mail:
lightnesstibet@163.com), and Department of Electrical Engineering, Wright State University (zhiqiang.wu@wright.edu)}
}

\markboth{}
{Zhang \MakeLowercase{\textit{et al.}}: Manifold Optimization Based Multi-user Rate Maximization Aided by Intelligent Reflecting Surface}

\maketitle

\begin{abstract}
In this work, two problems associated with a downlink multi-user system are considered  with the aid of intelligent reflecting surface (IRS): weighted sum-rate maximization and  weighted minimal-rate maximization.
For the first problem, a novel DOuble Manifold ALternating Optimization (DOMALO) algorithm is proposed by exploiting the matrix manifold theory and introducing  the beamforming matrix and reflection vector using complex sphere manifold and complex oblique manifold, respectively, which incorporate the inherent geometrical structure and the required  constraint.
A smooth double manifold alternating optimization (S-DOMALO) algorithm is then developed based on  the Dinkelbach-type algorithm and smooth exponential penalty function for the second problem.
 Finally, possible cooperative beamforming gain between IRSs and the IRS phase shift with limited resolution is studied, providing a reference for practical implementation. Numerical results show that our proposed algorithms can significantly outperform the benchmark schemes.
\end{abstract}


\IEEEpeerreviewmaketitle

\section{Introduction}

Intelligent reflecting surface (IRS) has been regarded as a promising solution for enhancing  the 6G-based Internet of Things (IoT) networks\cite{towards,9509294}. IRS is a two-dimension (2D) meta-surface composed of reconfigurable and near-passive reflecting elements with less energy drive, each of which can independently adjust the amplitude and phase of the incident signal. This advantage makes IRS widely applied to improve the signal propagation conditions and  enhance the spectral efficiency and energy efficiency of wireless communication systems and the data offloading rates for IoT systems \cite{MIMO1,9509294}.
By optimizing the phase shift of each IRS reflecting element and the beamforming matrix of the base station (BS) according to the perfect channel state information (CSI), we can reduce the transmission power \cite{li2019joint,wu2020joint}, extend the wireless coverage \cite{cao2020intelligent,peng2020multiuser}, improve physical layer security \cite{cui2019secure,dong2020secure,9612585}, boost radar-communication system performance \cite{9416177,9264225}, enhance the efficiency
of both wireless information transfer and wireless
power transfer simultaneously \cite{9305278}, and improve the user rate \cite{cao2020intelligent,feng2020deep,8746155,9013288,chen2019sum,9531458,tang2020joint,kammoun2020asymptotic,xie2020max}.

%In addition, IRS has also been combined with other communication technologies to improve the user rate \cite{yang2019irs,zhao2020two,yang2020irs,yang2020intelligent}. In \cite{yang2019irs}, authors consider the rate maximization problem in orthogonal frequency division multiplexing (OFDM) system and solve it with \textcolor{red}{full name of SDR (SDR) and (SCA)} methods. The authors in \cite{zhao2020two} use the weighted minimum mean-squared error (WMMSE) and SCA methods to IRS-aided OFDM system. In \cite{yang2020irs}, the authors consider an IRS-aided system employing orthogonal frequency division multiple access (OFDMA) and maximize the minimum rate by leveraging alternating optimization and SCA methods. In \cite{yang2020intelligent}, the authors investigate non-orthogonal-multiple-access (NOMA) and optimize the rate performance and ensure user fairness through block coordinate descent (BCD) and SDR techniques.

In particular, for user rate improvement, the authors in \cite{feng2020deep} considered a single-user scenario. In \cite{8746155}, the authors utilized statistical CSI to increase the ergodic user rate in the single-user scenario, which improves the communication system's robustness. In contrast, a multi-user scenario is considered in \cite{9013288}, with one IRS employed in the system. By utilizing the additional degrees of freedom in frequency, power, and code domains, IRS can also be combined with other techniques, such as orthogonal frequency division multiplexing (OFDM) \cite{zhao2020two,yang2020irs,9497353} or non-orthogonal multiple access (NOMA) \cite{yang2020intelligent}, to further improve the user rate.
In this paper, we consider a more generalized model of a multi-user multi-IRS system, where the single-user single-IRS problem can be treated as a special case. Naturally, the user rate optimization problem for such a system becomes more complicated.

Many iterative optimization methods have been successfully applied to capacity improving problems in IRS-aided systems, such as the local search and cross-entropy (CE) based algorithm for sum-rate optimization \cite{chen2019sum}.
These methods usually rely on greedy search with high computational complexity. The authors in \cite{9531458}, exploiting CE-based framework, proposed a machine learning-aided algorithmic method to solve IRS-aided communication system design problems, including weighted sum-rate maximization.
In \cite{cao2020intelligent}, a distributed IRS deployment scheme was proposed to avoid low-rank BS-IRS mmWave channel, and the problem was then transformed into a semi-definite programming (SDP) form. An
%algorithm to solve this non-convex problem by using techniques of
alternating optimization and semi-definite relaxation (SDR) algorithm was developed  to solve the max-min problem in \cite{tang2020joint}. In contrast, in \cite{kammoun2020asymptotic}, a projected gradient ascent algorithm was introduced to solve this problem. In \cite{xie2020max}, three different alternating-optimization-based solutions were provided for the max-min problem, including bisection search, successive convex approximation (SCA), and sub-gradient projection. In \cite{9362274}, the authors reconsidered the max-min problem from utilizing inter-IRS channel and proved that the multi-IRS scheme performs better than the single-IRS scheme. The problem was solved by the sub-optimal closed-form expressions, the SDR, and the bisection methods. In \cite{9417491}, using two SCA-based algorithms and a greedy algorithm, the authors optimized BS beamforming vectors, IRS switch vector, and reflection coefficients of IRSs iteratively to maximize the sum-rate under minimum user rate, unit-modulus of the reflection coefficients, and transmission power constraints. %\textcolor{red}{
However, by simply applying these non-convex optimization methods to solve the rate maximization problem, the signal model's inherent structure is ignored in the process, which may provide an opportunity to reduce the computational complexity and improve the performance further.
%}

%However, these techniques mentioned above are all traditional methods for solving non-convex optimization problems.  Thus, we are motivated to explore non-convex optimization methods to maximize the overall user sum-rate and minimal-rate with low computational complexity and high performance.

%\textbf{Why manifold optimization}:
In recent years,  significant development in Riemannian manifold optimization for various matrix manifolds has been made. In \cite{li2020manifold}, for the joint design of transmit waveform and receive filter in a multiple input multiple output space-time adaptive processing (MIMO-STAP) radar, the constraint was transformed into a smooth and compact manifold to reduce the computational cost. A channel estimation algorithm was proposed in the IRS-aided MIMO system by leveraging the fixed-rank manifold \cite{9322519}. In \cite{hong2020semi} a sphere manifold was used for data detection.
Moreover, the authors pioneered the use of manifolds to solve the optimization problem of the IRS-aided communication system and proved that the use of manifolds has the advantage of being spectrally efficient and computationally efficient \cite{8855810}.
Therefore, by considering the geometrical structures and low-dimension feature of the manifold, we could solve some non-convex problems with high efficiency when the feasible sets have manifold features, especially in high dimensions \cite{douik2019manifold}.

This paper will revisit the IRS-aided user rate improvement problem from a manifold optimization perspective. The main contributions are summarized as follows:

\begin{itemize}
\item Two schemes for multi-user multi-IRS capacity optimization are proposed: the weighted sum-rate optimization and the weighted minimal-rate optimization. They follow a unified algorithm framework that jointly optimizes the BS beamforming matrix and the IRS reflection vector, subject to the maximum transmission power and unit modulus constraints.
\item To solve the non-convex weighted sum-rate maximization problem, a double manifold alternating optimization (DOMALO) algorithm is developed  in IRS-aided %\textcolor{red}{multiple-input single-output (MISO)}
systems with inter-user interference. Specifically, the structural information of the constraint is exploited, and the beamforming matrix and IRS reflection vector are replaced by the complex sphere and complex oblique manifolds, respectively. Via alternative iteration using a geometric conjugate gradient, the sum-rate is improved significantly.
\item To guarantee an achievable rate for all users, the max-min fairness problem is investigated in the IRS-aided multi-user system, and a manifold optimization method named smooth double manifold alternating optimization (S-DOMALO) is proposed, which  introduces the Dinkelbach-type algorithm and smooth exponential penalty function to improve the weighted minimal-user rate.
\item To provide further insight into the performance of the IRS-aided communication system under general conditions, the impact of the reflection coefficients quantization and the possible cooperative beamforming gain between IRSs is considered.

\item  The preconditions for the inter-IRS channels to affect the system performance are also analyzed. As demonstrated by simulation results, the
proposed schemes outperform some existing approaches regardless of the numbers of BS transmit antennas, IRS elements, users, and SNR levels in the IRS-aided communication system.
\end{itemize}

In our earlier conference paper \cite{9580434}, we presented the basic idea of using manifold optimization for maximizing weight sum-rate and provided some preliminary results. The difference between this paper and \cite{9580434} can be boiled down to four aspects. First, more details are provided to the proposed DOMALO algorithm, such as by  adding convergence and complexity analysis. Second, this paper studies both the maximizing weighted sum-rate and the weighted minimal-rate problems based on a unified manifold optimization framework, while  the previous paper only studied the former problem. Third, this paper considers the possible cooperative beamforming gain between IRSs due to the existence of  inter-IRS channels. Finally, more extensive  simulation results and analyses are provided in this paper.

The remainder of this paper is organized as follows. Section \ref{System Model} introduces the system model of the IRS-aided multi-user system. A double manifold alternating optimization algorithm is proposed to solve the weighted sum-rate problem in Section \ref{Maximize Sum-Rate}. In Section \ref{Maximize Minimal-Rate}, the max-min fairness issue is considered, and the weighted minimal-rate problem is solved through the manifold optimization. Simulation results are presented in Section \ref{Simulation Results} and conclusions are drawn  in Section \ref{Conclusion}.

\textit{Notations}:
%Scalars are denoted by italic letters, vectors and matrices are denoted by bold-face lower-case and upper-case letters,
Scalar, vector, matrice, and manifold are denoted by italic letter $a$, lower-case boldface letter $\mathbf{a}$, upper-case boldface letter $\mathbf{A}$ and calligraphy letter $\mathcal{A}$, respectively. $ \mathbb{C}^{n\times p} $ denotes the space of $ n\times p $ complex-valued matrices.
% \textcolor{red}{ $\mathrm{vec}\left(\cdot \right)$ vectorize? $\mathrm{{tr}(\dot)$ is the inverse operation of}
$\mathrm{vec}(\mathbf{A})$ and $\mathrm{tr}(\mathbf{A})$ denote the vectorization and the trace operation, respectively.
$ \mathrm{diag}\left(\mathbf{a} \right)  $ denotes the diagonal operation and %$ \mathrm{diag}\left(\cdot \right)  $.
$ \mathrm{ddiag}\left(\mathbf{A} \right)  $ is a operation that sets all off-diagonal entries of a matrix to zero. $ \mathbb{E}\{a\} $  is the expectation operator. $ \otimes $ is the Kronecker product. $ \mathbf{A} ^{*} $, $ \mathbf{A}^{T} $ and $\mathbf{A}^{H}$ denote conjugate, transpose, and conjugate transpose operations, respectively. $\mathbf{1}_M$ denotes $M\times 1$ column vector of all ones. %\textcolor{red}{All vectors and matrices are...., Manifold ...}

\section{System Model}\label{System Model}

\begin{figure}[htbp]
	\centerline{\includegraphics[width=0.45\textwidth]{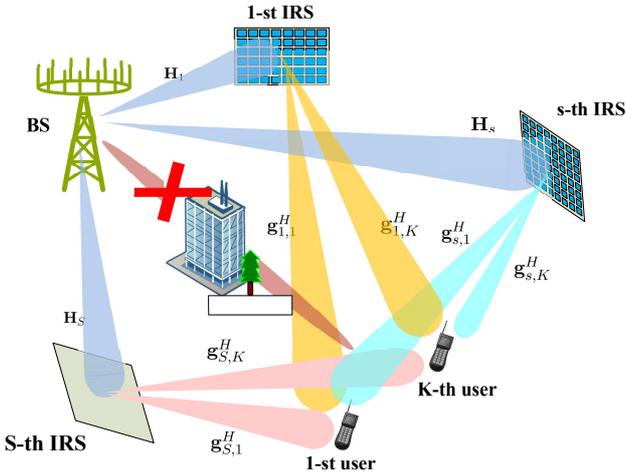}}
	\caption{The down-link channels in the IRS-aided multi-user system.}
	\label{v}
\end{figure}

Consider an IRS-aided communication system with $ K $ users equipped with a single antenna,  a BS with $ N $ antennas, and $ S $ IRSs each consisting of $ M $ elements, as shown in Fig. \ref{v}.
Assume that all of the IRSs are exactly the same\footnote{Note that the condition of the same type is not necessary, and the later analysis is valid for IRSs with arbitrary sizes. The assumption on the same types of IRS is just for the convenience of explanation.} and there are no direct links/channels between the BS and users, or the links are so weak that they can be ignored. The reflection matrix of IRS, which is diagonal and composed of the reflection coefficients of all elements, is denoted as $\mathbf{\Phi}_{s}=\mathrm{diag}(\gamma_1e^{j\phi _{1}},\gamma_2e^{j\phi _{2}},\dots,\gamma_Me^{j\phi _{M}} )$, where $\gamma_m$ and $\phi_m$ denotes the reflection amplitude and the $m$-th element's phase shift, respectively. %\textcolor{red}{it's suggested to modidy $\theta$ to be $\phi$ here, to consists with $\Phi$}

The complex transmitted signal from the BS is given by
\begin{equation}
	\mathbf{x}=\sum_{k=1}^{K}\mathbf{v}_{k}s_{k},
	\label{eq_TX}
\end{equation}
where $ \mathbf{v}_{k} $ denotes  the beamforming vector from the BS to the $ k $-th user, and $ s_{k} \sim \mathcal{CN}(0,1)$ is the transmitted symbols.
The received signal $ y_k $ at user $ k $, without considering the inter-IRS channels, is
\begin{equation}\label{receive_signal}
	y_{k}=\left(\sum_{s=1}^{S}\mathbf{g}_{s,k}^{H}\mathbf{\Phi }_{s}\mathbf{H}_{s}\right)\mathbf{x}+u_{k},
\end{equation}
where $\mathbf{H}_{s}\in \mathbb{C}^{M\times N} $ denotes the channel between the BS and the $s$-th IRS, $\mathbf{g}_{s,k}^{H}\in \mathbb{C}^{1\times M} $ denotes the channel between the $s$-th IRS and the $k$-th user, and $ u_{k}\sim \mathcal{CN}(0,\sigma_k^2) $ is the additive white Gaussian noise (AWGN) for the $ k $-th user.

%\subsection{Impact of inter-IRS channels}\label{inter-IRS channels}
%This subsection mainly discusses a double-IRS aided communication system with one inter-IRS channel, The similar goes for the multi-IRS scenario.

%Based on the above system model,
Therefore, the achievable rate for the $ k $-th user, $k=1, \dots, K$, is
\begin{equation}\label{Rk}
	\mathcal{R}_{k}=\log _{2}(1+r_{k}),
\end{equation}
where
%$ r_{k} $ in (\ref{Rk}) is the Signal to Interference plus Noise Ratio (SINR) of $ k $-th user and can be expressed as
\begin{equation}
r_{k}=\frac{|\mathbf{g}_{k}^{H}\mathbf{\Phi }\mathbf{H}\mathbf{v}_{k}|^{2}}{\sum_{j\neq k}^{K}|\mathbf{g}_{k}^{H}\mathbf{\Phi }\mathbf{H}\mathbf{v}_{j}|^{2}+\sigma _{k}^{2}},
\label{eq_SINR}
\end{equation}
%where $ \mathbf{g}_{k}^{H} $, $ \mathbf{H} $ and $ \mathbf{\Phi } $ is:
with $\mathbf{g}_{k}^{H}=\left[ \mathbf{g}_{1,k}^{H},\mathbf{g}_{2,k}^{H},\dots,\mathbf{g}_{S,k}^{H}\right]$, $\mathbf{H}=\left[ \mathbf{H}_{1}^T, \mathbf{H}_{2}^T,\dots,\mathbf{H}_{S}^T\right] ^{T}$, and $\mathbf{\Phi }=\mathrm{diag}(\mathbf{\Phi }_{1},  \mathbf{\Phi }_{2}, \dots, \mathbf{\Phi }_{S})=\mathrm{diag}(\gamma_1e^{j\phi _{1}},\gamma_2e^{j\phi _{2}},\dots,\gamma_{SM}e^{j\phi _{SM}} )$. For the convenience of discussion, suppose the reflection  amplitudes are the same, denoted as $\gamma$.

The inter-IRS channels are passive-to-passive channels, and the path-loss is very high because it is in the form of product rather than addition \cite{bjornson2021reconfigurable,9497353}. %Therefore, it seems not cost-effective and unnecessary to consider them.
%\footnote{ It is worth mentioning that although the algorithms proposed in this paper do not consider inter-IRS channels, they can be generalized to the general case with inter-IRS channels. The specific details are described in the next section.}
Considering inter-IRS channels, and taking  double-IRS situation as an example, \eqref{receive_signal} becomes
%Assuming that IRS1 is close to BS and IRS2 is close to users, we need to consider only one dual cascade channel: BS-IRS1-IRS2-users. Similar to (\ref{receive_signal}), considering inter-IRS channel, the received signal can be expressed as
\begin{equation}
	\breve{y}_{k}=\left(\sum_{s=1}^{2}\mathbf{g}_{s,k}^{H}\mathbf{\Phi }_{s}\mathbf{H}_{s}+\mathbf{g}_{2,k}^{H}\mathbf{\Phi }_{2}\mathbf{\Lambda}_{1,2}\mathbf{\Phi }_{1}\mathbf{H}_{1}\right)\mathbf{x}+u_{k},
\end{equation}
where $ \mathbf{\Lambda}_{1,2}\in \mathbb{C}^{M\times M}$ denotes the channel between the first ($s=1$) IRS and the second ($s=2$) IRS. Then,
\eqref{eq_SINR} becomes
\begin{equation}
r_{k}=\frac{\left|\left(\sum_{s=1}^{2}\mathbf{g}_{s,k}^{H}\mathbf{\Phi }_{s}\mathbf{H}_{s}+\mathbf{g}_{2,k}^{H}\mathbf{\Phi }_{2}\mathbf{\Lambda}_{1,2}\mathbf{\Phi }_{1}\mathbf{H}_{1}\right)\mathbf{v}_{k}\right|^{2}}{\sum_{j\neq k}^{K}\left|\left(\sum_{s=1}^{2}\mathbf{g}_{s,k}^{H}\mathbf{\Phi }_{s}\mathbf{H}_{s}+\mathbf{g}_{2,k}^{H}\mathbf{\Phi }_{2}\mathbf{\Lambda}_{1,2}\mathbf{\Phi }_{1}\mathbf{H}_{1}\right)\mathbf{v}_{j}\right|^{2}+\sigma _{k}^{2}},
\label{eq_SINR2}
\end{equation}
accordingly. A similar result goes for the multi-IRS scenario.

%This actually unifies the distributed IRS model and the centralized IRS model used in many existing literature.
% \begin{equation}
% 	\mathbf{\Phi }=\begin{bmatrix}
% 		\mathbf{\Phi }_{1} & \cdots &0 \\
% 		\vdots & \ddots &\vdots\\
% 		0 & \cdots  & \mathbf{\Phi }_{S}
% 	\end{bmatrix}.
% \end{equation}

To improve the user rate in \eqref{Rk} by jointly optimizing the beamforming vectors $\mathbf{v}_k$ at the BS and reflection matrix $\mathbf{\Phi}$ at the IRS,
we have to turn the problem into the convex form such that the CVX toolbox \cite{cvx} or other similar optimization tools can be applied.
Note that for situations with multiple users and IRSs,
solving  the problem is usually of high computational complexity.
In the following, the inherent geometrical structure in our considered system is examined and an innovative manifold optimization scheme is proposed to maximize the sum-rate or the minimal-rate.

\section{Maximizing Weighted Sum-Rate}\label{Maximize Sum-Rate}
In this section, a manifold optimization based algorithm is developed by jointly optimizing the beamforming matrix and reflection matrix to find the suboptimal solution to the weighted sum-rate problem.
\subsection{Problem Formulation}\label{PF}

Considering the BS transmission power constraint and the IRS reflection coefficient constraint, the weight sum-rate maximization optimization problem ($\mathrm{P1}$) for multiple users can be formulated as
\begin{equation}\label{P1}
\begin{aligned}
\left( \mathrm{P1}\right) :&\;\underset{\mathbf{V},\mathbf{\Phi} }{\mathrm{max}}\: f_{1}(\mathbf{V},\mathbf{\Phi} )=\sum_{k=1}^{K}\omega _{k}\mathcal{R}_{k}
\\&\mathrm{s.t.}\; \mathrm{tr}(\mathbf{VV}^{H})\leq P
\\&\phi_{i}\in \mathcal{F},i=1,2,\dots,SM,
\end{aligned}
\end{equation}
where $\mathbf{V}=[\mathbf{v}_{1},\mathbf{v}_{2},\cdots,\mathbf{v}_{K}] $ denotes the beamforming matrix according to \eqref{eq_TX}, the weight $ \omega _{k} $ is the required service priority of the $ k $-th user and $ P $ is the maximum transmission power at BS. In addition, $ \mathcal{F} $ denotes the feasible set of the reflection coefficients, which has two different options. One has infinite resolution under ideal conditions, and the other is of finite resolution controlled by a $ Q $-level quantizer. %\textcolor{red}{The details will be explained in \ref{E}.(no use)}

\subsection{Reformulation of the Original Problem without Considering the Inter-IRS Channel}\label{A}
In order to deal with the sum-log problem, we decouple and divide the problem into three sub-problems as in \cite{shen2018fractional,9459505}. By introducing auxiliary variables $\bm{\zeta}\in \mathbb{C}^{1\times K} $ with $\bm{\zeta}=\left[\zeta_{1},\cdots,\zeta_{k},\cdots,\zeta_{K} \right]$, the original problem $( \mathrm{P1}) $ is reformulated as
\begin{equation}
\begin{split}
\left( \mathrm{P2}\right): &\;\underset{\mathbf{V},\mathbf{\Phi},\bm{\zeta}}{\mathrm{max}} \;f_{2}(\mathbf{V},\mathbf{\Phi},\bm{\zeta})=\frac{1}{\ln 2}\left(\sum_{k=1}^{K}\omega _{k}\ln(1+\mathbf{\zeta }_{k})\right. \\& \left.+\sum_{k=1}^{K}\left(-\omega _{k}\mathbf{\zeta }_{k}+\frac{\omega _{k}(1+\mathbf{\zeta }_{k})r_{k}}{1+r_{k}}\right)\right)
\\& \mathrm{s.t.}\; \mathrm{tr}(\mathbf{VV}^{H})\leq P
\\&\phi_{i}\in \mathcal{F},i=1,2,\dots,SM.
\end{split}
\label{eq_P2}
\end{equation}

Then, we alternately optimize $ \mathbf{V}$, $ \mathbf{\Phi} $  and $ \bm{\zeta} $ by fixing the other two variables and optimize the remaining one until convergence of the objective function is achieved.
\subsection{Fix $ \mathbf{V},\mathbf{\Phi} $ and Optimize $ \bm{\zeta} $}\label{B}

During each iteration, the auxiliary variable $ \bm{\zeta} $ is updated firstly according to the current value of $ \mathbf{V} $ and $ \mathbf{\Phi} $, through setting $\partial f_{2}(\mathbf{V},\mathbf{\Phi},\bm{\zeta})/\partial \bm{\zeta}  $ to zero. The updated $ \bm{\zeta} $ is expressed as
\begin{equation}\label{zeta}
	{\mathbf{\zeta }}_{k}=r_{k}.
\end{equation}

Since $f_{2}$ is a concave differentiable function over $\bm{\zeta}$ when $ \mathbf{V}$ and $\mathbf{\Phi} $ are fixed,  we can recover $f_{1}$ exactly through substituting (\ref{zeta}) back in $f_{2}$. Therefore, ($\mathrm{P2}$) and ($\mathrm{P1}$) are equivalent. In addition, notice that only one term of the objective function in $ (\mathrm{P2}) $ is related to $ \mathbf{V}$  and $\mathbf{\Phi} $. So when optimizing $ \mathbf{V}$  and $\mathbf{\Phi} $, the objective function can be further simplified as
\begin{equation}
	\begin{aligned}
		\left( \mathrm{P3}\right):&\; 	\underset{\mathbf{V},\mathbf{\Phi }}{\mathrm{max}}\; f_{3}(\mathbf{V},\mathbf{\Phi })\\&=\sum_{k=1}^{K}\frac{\omega _{k}(1+\zeta _{k})r_{k}}{1+r_{k}}
		\\&=\sum_{k=1}^{K}\frac{\omega _{k}(1+\zeta _{k})|(\mathbf{g}_{k}^{H}\mathbf{\Phi }\mathbf{H})\mathbf{v}_{k}|^{2}}{\sum_{j=1}^{K}|(\mathbf{g}_{k}^{H}\mathbf{\Phi }\mathbf{H})\mathbf{v}_{j}|^{2}+\sigma _{k}^{2}}
		\\& \mathrm{s.t.}\; \mathrm{tr}(\mathbf{VV}^{H})\leq P
		\\&\phi_{i}\in \mathcal{F},i=1,2,\dots,SM.
	\end{aligned}
\end{equation}

\subsection{Fix  $ \bm{\zeta} $ , $\mathbf{\Phi} $ and Optimize $ \mathbf{V} $}\label{C}

Given fixed $\mathbf{\Phi} $, $ (\mathrm{P3}) $ is equivalent to
\begin{equation}
		\begin{aligned}
			\left( \mathrm{P3'}\right):&\;	\underset{\mathbf{V}}{\mathrm{max}}\; f_{3}(\mathbf{V})
			\\&=\sum_{k=1}^{K}\frac{\omega _{k}(1+\zeta _{k})|(\mathbf{g}_{k}^{H}\mathbf{\Phi }\mathbf{H})\mathbf{v}_{k}|^{2}}{\sum_{j=1}^{K}|(\mathbf{g}_{k}^{H}\mathbf{\Phi }\mathbf{H})\mathbf{v}_{j}|^{2}+\sigma _{k}^{2}}
			\\& \mathrm{s.t.}\; \mathrm{tr}(\mathbf{VV}^{H})\leq P,
		\end{aligned}
\end{equation}

Without loss of generality, we normalize the power to get $ \mathrm{tr}(\mathbf{VV}^{H})\leq 1 $ and define $ \mathbf{\hat{V}}=[\mathbf{\hat{v}}_{1},\mathbf{\hat{v}}_{2},\cdots,\mathbf{\hat{v}}_{K}]  $ satisfying $ \mathrm{tr}(\mathbf{\hat{V}\hat{V}}^{H})=\mathrm{tr}(\mathbf{VV}^{H})+||\mathbf{\varkappa}||_2^2= 1 $, where $ \mathbf{\hat{v}}_{k}=[\mathbf{v}_{k}^T,\varkappa_k]^T $ and $ \mathbf{\varkappa}=[\varkappa_1,\varkappa_2,\cdots,\varkappa_K] $.

As the Frobenius norm is equal to $1$, we then define a manifold as $ \mathcal{M}_{1}=\left\lbrace  \mathbf{\hat{V}}\in \mathbb{C}^{(N+1)\times K}\big| \mathrm{tr}(\mathbf{\hat{V}\hat{V}}^{H})= 1 \right\rbrace $, such that $ (\mathrm{P3'}) $ is equivalent to the unconstrained optimization on $  \mathcal{M}_{1} $, that is
\begin{equation}
\begin{aligned}
\left( \mathrm{P3''}\right):&\;	\underset{\mathbf{\hat{V}}\in \mathcal{M}_{1}}{\mathrm{max}}\; f_{3}(\mathbf{\hat{V}})
\\&=\sum_{k=1}^{K}\frac{\omega _{k}(1+\zeta _{k})|(\mathbf{g}_{k}^{H}\mathbf{\Phi }\mathbf{\hat{H}})\mathbf{\hat{v}}_{k}|^{2}}{\sum_{j=1}^{K}|(\mathbf{g}_{k}^{H}\mathbf{\Phi }\mathbf{\hat{H}})\mathbf{\hat{v}}_{j}|^{2}+\sigma _{k}^{2}},
\end{aligned}
\end{equation}
where $ \mathbf{\hat{H}}=\sqrt{P}\left[ \mathbf{H},\mathbf{0}\right]  $.

%The method of this kind of manifold optimization problem is given in \cite{al}.

% We modify geometric conjugate gradient(GCG) algorithm mentioned in \cite{al} to slove the subproblem of the beamforming at BS. We will introduce the two core concepts in \textbf{Algorithm 1} below.
Recalling the main idea of geometric conjugate gradient (GCG) \cite{al}, we introduce the geometric conjugate gradient on the manifold algorithm to solve the above subproblem, as shown in \textbf{Algorithm 1}.
%\textcolor{red}{$i$ can not be the iteration index here, try to use $t$ and replace $t$ in (35) . The same as Algorithm 2 and 3.}
 \begin{itemize}
    \item Retraction.
    The next iteration point can be determined by the step size and search direction for the conjugate gradient algorithm studied in Euclidean space. But if the selectable variables constitute a Riemannian sub-manifold, it is very likely that the next iteration point does not fall into the manifold. To this end, we need to make retractions for specific manifolds to ensure that the next iteration point is still on the manifold. In \textbf{Algorithm 1}, retraction is denoted as $ \mathcal{R} $. For example, the retraction of $ \alpha _{t}\eta _{t} $ ($ \alpha _{t} $ is the Armijo step size) %\textcolor{red}{and $\mathbf{x}_i$ is the transmit signal, $i$ is the index of IRS elements}
    at point $ \mathbf{p}_{t} $ on the manifold $ \mathcal{M} $ can be expressed as $ \mathcal{R}_{\mathbf{p}_{t}}(\alpha _{t}\eta _{t}) $.
    \item Vector transport.
    If the tangent plane at point $ \mathbf{p}_{t} $ is defined as $ T\mathcal{M}_{\mathbf{p}_{t}} $, we know that the Riemannian gradient $\mathrm{Rgrad} F (\mathbf{p}_{t+1})$ at $ \mathbf{p}_{t+1}   $ belongs to $ T\mathcal{M}_{\mathbf{p}_{t+1}} $.  However, $\eta _{t}$ indicates that  the search direction at $ \mathbf{p}_{t} $ belongs to $T\mathcal{M}_{\mathbf{p}_{t}}$. Since $ \mathrm{Rgrad} F (\mathbf{p}_{t+1}) $ and $ \eta _{t} $ are not in the same tangent plane, they cannot be added directly. To this end, we need to transport $ \eta _{t} $ to $T\mathcal{M}_{\mathbf{p}_{t+1}} $. In \textbf{Algorithm 1}, the operator
 of vector transport is denoted as $ \mathcal{T} $.  For example, vector transport of $ \eta _{t} $ to $T\mathcal{M}_{\mathbf{p}_{t+1}} $ by combining $ \mathcal{R}_{\mathbf{p}_{t}}(\alpha _{t}\eta _{t}) $  can be expressed as $\mathcal{T}_{\alpha _{t}\eta _{t}}(\eta _{t}) $.
\end{itemize}

\begin{figure}[htbp]
  	\centerline{\includegraphics[width=0.45\textwidth]{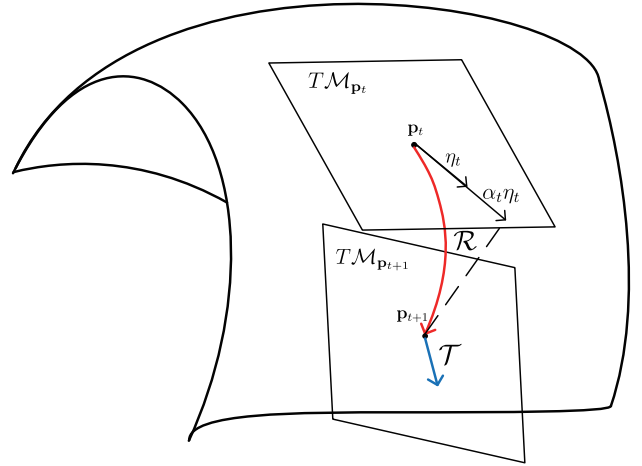}}
  	\caption{Geometric interpretation of the two key concepts at the  $t+1$-th iteration.}
  	\label{g}
  \end{figure}

 The two key concepts are geometrically illustrated in Fig. \ref{g}, where the red line represents retraction, and the blue line represents vector transport.

\begin{algorithm}[ht]
	\caption{Geometric Conjugate Gradient on Manifold Algorithm}
	\begin{algorithmic}[1]
		\REQUIRE Retraction $ \mathcal{R} $ on $ \mathcal{M} $ and vector transport $ \mathcal{T} $ on $ \mathcal{M} $, the cost function $ F  $;
		 \STATE Initialize  point $ \mathbf{p}_{0}\in \mathcal{M} $;
		 \STATE Initialize search direction $\eta _{0}=\mathrm{Rgrad} F(\mathbf{p}_{0}) $;
		 \STATE Iteration counter $ t=0 $;
		\REPEAT
		\STATE Compute Armijo step size $\alpha_{t}>0 $;
		\STATE Update point  $\mathbf{p}_{t+1}=\mathcal{R}_{\mathbf{p}_{t}}(\alpha _{t}\eta _{t}) $;
	    \STATE Compute correction parameter $ \beta _{t+1}=\frac{\left \langle \mathrm{Rgrad}F(\mathbf{p}_{t+1}), \mathrm{Rgrad}F(\mathbf{p}_{t+1}) \right \rangle}{\left \langle \mathrm{Rgrad}F(\mathbf{p}_{t}),\mathrm{Rgrad}F(\mathbf{p}_{t}) \right \rangle} $;
    	\STATE Update search direction $ \mathbf{\eta }_{t+1}=\mathrm{Rgrad} F(\mathbf{p}_{t+1})+\beta _{t+1}\mathcal{T}_{\alpha _{t}\eta _{t}}(\eta _{t}) $;
    		\STATE $t=t+1 $;
    	\UNTIL Convergence;
	    \ENSURE Converged point $\mathbf{p}_{t}  $.
	\end{algorithmic}
\end{algorithm}

As the GCG algorithm uses first-order information of the cost function, the critical step is to calculate the Riemannian gradient of the objective function at the current point.
By calculating the Euclidean gradient of the cost function and projecting the Euclidean gradient onto the tangent space \cite{boumal2020introduction}, we obtain the Riemannian gradient of the cost function.
In addition, as the variable to be optimized in the cost function equation is $\left\lbrace \hat{\mathbf{v}}_{k}\right\rbrace  $, which is a certain column of $ \hat{\mathbf{V}} $, and the constraint is about $ \hat{\mathbf{V}} $, we cannot use the GCG algorithm directly. Therefore we construct an index matrix, which is a $ k $-order identity matrix $\mathbf{E}_{k} $, so that any column of $ \hat{\mathbf{V}} $ can be represented using the index matrix. If the $ i $-th column of $\mathbf{E}_{k} $ is expressed as $\mathbf{E}_{ki} $, then $ (\mathrm{P3''}) $ is rewritten as
\begin{equation}\label{P4}
\begin{aligned}
	\left( \mathrm{P4}\right):&\underset{\hat{\mathbf{V}}\in\mathcal{M}_{1}}{\mathrm{max}}\; f_{4}(\hat{\mathbf{V}})=\\&\sum_{k=1}^{K}\frac{\tilde{\zeta} _{k}|(\mathbf{g}_{k}^{H}\mathbf{\Phi }\hat{\mathbf{H}})\hat{\mathbf{V}}\mathbf{E}_{kk}|^{2}}{\sum_{j=1}^{K}|(\mathbf{g}_{k}^{H}\mathbf{\Phi }\hat{\mathbf{H}})\hat{\mathbf{V}}\mathbf{E}_{kj}|^{2}+\sigma _{k}^{2}},
\end{aligned}
\end{equation}
where $\tilde{\zeta }_{k}$ is equal to $ \omega _{k}(1+\zeta _{k}) $.

In order to  simplify  the expression of the Euclidean gradient of the objective function for $ \hat{\mathbf{V}} $ in $ (\mathrm{P4}) $, define $ \tilde{\mathbf{h}}_{k}^{H}=\mathbf{g}_{k}^{H}\mathbf{\Phi }\hat{\mathbf{H}} $, and then the Euclidean gradient becomes (\ref{EV}), shown at the top of next page.
\begin{figure*}
\begin{equation}\label{EV}
\mathrm{Egrad}\;f_{4}(\hat{\mathbf{V}})=\frac{\partial f_{4}(\hat{\mathbf{V}})}{\partial \hat{\mathbf{V}}}=
\sum_{k=1}^{K}\tilde{\zeta }_{k}\cdot \left( \frac{2\tilde{\mathbf{h}}_{k}^{H}\hat{\mathbf{V}}\mathbf{E}_{kk}\tilde{\mathbf{h}}_{k}\mathbf{E}_{kk}^{H}}{\sum_{j=1}^{K}|\tilde{\mathbf{h}}_{k}^{H}\hat{\mathbf{V}}\mathbf{E}_{kj}|^{2}+\sigma _{k}^{2}}
-2\sum_{i=1}^{K}\frac{|\tilde{\mathbf{h}}_{k}^{H}\hat{\mathbf{V}}\mathbf{E}_{kk}|^{2}\tilde{\mathbf{h}}_{k}^{H}\hat{\mathbf{V}}\mathbf{E}_{ki}\tilde{\mathbf{h}}_{k}\mathbf{E}_{ki}^{H}}{(\sum_{j=1}^{K}|\tilde{\mathbf{h}}_{k}^{H}\hat{\mathbf{V}}\mathbf{E}_{kj}|^{2}+\sigma _{k}^{2})^{2}}\right) .
\end{equation}
\hrule
\end{figure*}
The projection operator of the tangent plane at point $ \hat{\mathbf{V}} $ on $ \mathcal{M}_{1} $ is defined as
\begin{equation}\label{PV}
\mathbf{P}_{ \hat{\mathbf{V}}}\left( \mathbf{\Xi}\right) = \mathbf{\Xi} - \mathrm{tr}(\hat{\mathbf{V}}^H \mathbf{\Xi}) \hat{\mathbf{V}},
\end{equation}
where $ \mathbf{\Xi}\in \mathbb{C}^{(N+1)\times K} $ represents a matrix in the ambient space.
%\textcolor{red}{$\textbf{x}$ is the transmit signal in eq (1)}

According to (\ref{EV}) and (\ref{PV}), we have the Riemann gradient as
\begin{equation}\label{REV}
\mathrm{Rgrad}\;f_{4}(\hat{\mathbf{V}})=\mathbf{P}_{ \hat{\mathbf{V}}}\left( \mathrm{Egrad}\;f_{4}(\hat{\mathbf{V}})\right).
\end{equation}

Moreover, the retraction and vector transport on the complex sphere are
\begin{equation}\label{RV}
\mathcal{R}_{\hat{\mathbf{V}}_{t}}(\alpha _{t}\eta _{t})_{\mathcal{SP}} =\dfrac{\hat{\mathbf{V}}_{t}+\alpha _{t}\eta _{t}}{||\hat{\mathbf{V}}_{t}+\alpha _{t}\eta _{t}||}
\end{equation}
and
\begin{equation}\label{TV}
\mathcal{T}_{\alpha _{t}\eta _{t}}(\eta _{t})_{\mathcal{SP}}=\mathbf{P}_{ \mathcal{R}_{\hat{\mathbf{V}}_{t}}(\alpha _{t}\eta _{t})_{\mathcal{SP}}}\left( \eta _{t}\right),
\end{equation}
respectively, where the subscript $ (\cdot)_\mathcal{SP} $ indicates  that the retraction and vector transport is on the complex sphere manifold. %Similarly, $ (\cdot)_\mathcal{OB} $ in \ref{D} means that the retraction and vector transport is on the complex oblique manifold.

Finally, we can optimize $ \mathbf{\hat{V}} $ by applying \textbf{Algorithm 1} under fixed $ \bm{\zeta} $ and $\mathbf{\Phi} $ and obtain $ \mathbf{V} $ via $ \mathbf{V}=\mathbf{\hat{V}}(1:N,K) $.

\subsection{Fix  $ \bm{\zeta} $ , $ \mathbf{V} $ and Optimize $\mathbf{\Phi} $}\label{D}

%In this subsection, we adopt a similar manifold optimzization approach to the previous subsection.

In this sub-problem, assume that $ \gamma =1$, and the optimization algorithm only adjusts the phase of the reflection coefficient: $\phi _{1},\phi _{2},\dots,\phi _{SM}\in [0,2\pi)$. Define $\mathbf{u} = \left ( e^{j\phi _{1}},e^{j\phi _{2}},...,e^{j\phi _{SM}}\right ) ^{H} $ satisfying $\mathbf{u}^{*}=\mathrm{vec}(\mathbf{\Phi}) $
%. That is $\mathbf{u}=\left ( e^{j\theta _{1}},e^{j\theta _{2}},...,e^{j\theta _{SM}}\right ) ^{H}$. %Because $\mathbf{u} $ satisfies
and $ (\mathbf{u}\mathbf{u}^H)_{ii} = 1$, $i = 1, \dots, SM  $, such that $\mathbf{u} $ forms a complex oblique manifold $\mathcal{M}_{2}=\left\{\mathbf{u} \in \mathbb{C}^{S M \times 1} \mid\left(\mathbf{u u}^{H}\right)_{i i}=1, i=1: S M\right\}$. Naturally, we can derive $\mathbf{u}^{H}\mathrm{diag}(\mathbf{g}_{k}^{H})\mathbf{H}=\mathbf{g}_{k}^{H}\mathbf{\Phi }\mathbf{H} $. Then we rewrite $ (\mathrm{P3}) $ as
\begin{equation}\label{P5}
\begin{aligned}
\left( \mathrm{P5}\right):&\;\underset{\mathbf{u}\in \mathcal{M}_{2}}{\mathrm{max}} \;f_{5}(\mathbf{u})\\&=\sum_{k=1}^{K}\frac{\tilde{\zeta }_{k}|\bigl(\mathbf{u}^{H}\mathrm{diag}(\mathbf{g}_{k}^{H})\mathbf{H}\bigr )\mathbf{v}_{k}|^{2}}{\sum_{j=1}^{K}|\bigl(\mathbf{u}^{H}\mathrm{diag}(\mathbf{g}_{k}^{H})\mathbf{H}\bigr )\mathbf{v}_{j}|^{2}+\sigma _{k}^{2}}
.
\end{aligned}
\end{equation}

To solve the problem (P5), we adopt a similar manifold optimization approach to the previous subsection. The projection operator of the tangent plane at point $ \mathbf{u} $ on $ \mathcal{M}_{2} $ \cite{hu2020brief} can be characterized by
\begin{equation}\label{Pu}
\mathbf{P}_{ \mathbf{u}}\left( \bm{\xi}\right) = \bm{\xi} - \mathrm{ddiag}(\bm{\xi}\mathbf{u}^H)\mathbf{u},
\end{equation}
where $ \bm{\xi}\in \mathbb{C}^{SM\times 1} $ represents a vector in the ambient space.

Then, the Euclidean gradient of $ f_{5}(\mathbf{u}) $ is calculated as (\ref{Eu}), shown at the top of next page.
\begin{figure*}
\begin{equation}\label{Eu}
\begin{split}
&\mathrm{Egrad}\;f_{5}(\mathbf{u })=\frac{\partial f_{5}(\mathbf{u})}{\partial \mathbf{u}}=\\&\sum_{k=1}^{K}\tilde{\zeta }_{k}\cdot \left(\frac{2\mathbf{v}_{k}^{H}(\mathbf{H}^{H}\mathrm{diag}^{H}(\mathbf{g}_{k}^{H})\mathbf{u})\mathrm{diag}(\mathbf{g}_{k}^{H})\mathbf{H}\mathbf{v}_{k}}{\sum_{j=1}^{K}|(\mathbf{u}^{H}\mathrm{diag}(\mathbf{g}_{k}^{H})\mathbf{H})\mathbf{v}_{j}|^{2}+\sigma _{k}^{2}}-2\sum_{i=1}^{K}\frac{|(\mathbf{u}^{H}\mathrm{diag}(\mathbf{g}_{k}^{H})\mathbf{H})\mathbf{v}_{k}|^{2}\mathbf{v}_{i}^H(\mathbf{H}^{H}\mathrm{diag}^{H}(\mathbf{g}_{k}^{H})\mathbf{u})\mathrm{diag}(\mathbf{g}_{k}^{H})\mathbf{H}\mathbf{v}_{i}}{(\sum_{j=1}^{K}|(\mathbf{u}^{H}\mathrm{diag}(\mathbf{g}_{k}^{H})\mathbf{H})\mathbf{v}_{j}|^{2}+\sigma _{k}^{2})^{2}}\right).
\end{split}
\end{equation}
\hrule
\end{figure*}
We have the Riemann gradient as
\begin{equation}\label{REu}
\mathrm{Rgrad}f_{5}(\mathbf{u})=\mathbf{P}_{ \mathbf{u}}\left( \mathrm{Egrad}f_{5}(\mathbf{u})\right).
\end{equation}

The retraction and vector transport on complex oblique manifold are expressed as
\begin{equation}\label{Ru}
\begin{aligned}
&\mathcal{R}_{\mathbf{u}_{t}}(\alpha _{t}\eta _{t})_{\mathcal{OB}}=\\
&\mathrm{ddiag}\left(\left( \mathbf{u}_{t}+\alpha _{t}\eta _{t}\right)\left( \mathbf{u}_{t}+\alpha _{t}\eta _{t}\right)^H\right)^{-1/2}\left( \mathbf{u}_{t}+\alpha _{t}\eta _{t}\right)
\end{aligned}
\end{equation}
and
\begin{equation}\label{Tu}
\mathcal{T}_{\alpha _{t}\eta _{t}}(\eta _{t})_{\mathcal{OB}}=\mathbf{P}_{ \mathcal{R}_{\mathbf{u}_{t}}(\alpha _{t}\eta _{t})_{\mathcal{OB}}}\left( \eta _{t}\right),
\end{equation}
where $ (\cdot)_\mathcal{OB} $ indicates the retraction and vector transport  on the complex oblique manifold.
%\textcolor{red}{the function of $ \mathrm{normalize} $ is to make every element of the vector have a norm $1$.(where is normalize?)}

\begin{algorithm}[t]
	\caption{DOuble Manifold ALternating Optimization Algorithm}
	\begin{algorithmic}[1]
		\REQUIRE Channel state information about $ \mathbf{H} $ and $\left\lbrace  \mathbf{g}_{k}\right\rbrace $; user service priority  $ \omega _{k} $, $ \forall k $.
		\STATE Initialize $ \mathbf{\hat{V}}^{(0)}$ and  $ \mathbf{u}^{(0)} $;
		\FOR {$t=1,2,...$}
		\STATE Update $ \mathbf{\zeta}^{(t)} $ by (\ref{zeta});
		\STATE Update $ \mathbf{\hat{V}}^{(t)}$ to solve $(\mathrm{P4})$ using \textbf{Algorithm 1};
		\STATE Update $ \mathbf{u}^{(t)}$ by solve $(\mathrm{P5})$ using \textbf{Algorithm 1};
			\IF {$ \left| f_{1}(\mathbf{V}^{(t)},\mathbf{\Phi}^{(t)} )-f_{1}(\mathbf{V}^{(t-1)},\mathbf{\Phi}^{(t-1)} )\right| \leq\varepsilon  $ or $t>T_{max}$ }
				\BREAK
				\ENDIF
		\STATE $ t=t+1 $;
		\ENDFOR
		\ENSURE Optimized beamforming vecter $ \left\lbrace \mathbf{v}_{k}\right\rbrace $ for all users; optimized reflection matrix $ \mathbf{\mathbf{\Phi}} $; suboptimal value of $ f_{1}(\mathbf{V}^{(t)},\mathbf{\Phi}^{(t)} ) $.
	\end{algorithmic}
\end{algorithm}

%According to these steps, we can optimize $\mathbf{\Phi} $ by applying \textbf{Algorithm 1} under the condition of fixed $ \mathbf{\zeta } $ and $ \mathbf{V} $.

The entire optimization process can be summarized as \textbf{Algorithm 2}, namely the DOuble Manifold ALternating Optimization algorithm (DOMALO).

\subsection{Considering the Inter-IRS Channel Case}\label{InterIRS}
Considering the inter-IRS channels case, as shown in \eqref{eq_SINR2}, equation  \eqref{zeta} in subsection \ref{B} still hold.
But in this case, the reflection matrix of IRS needs to be optimized separately.
Similar to the definition of $\mathbf{u}$ in \ref{D}, we have $\mathbf{u}_1^{*}=\mathrm{vec}(\mathbf{\Phi}_1) $ and $\mathbf{u}_2^{*}=\mathrm{vec}(\mathbf{\Phi}_2) $.
The sub-problems $f_4(\hat{\mathbf{V}})$ and $f_5(\mathbf{u})$ are adjusted to
{\small\begin{equation}\label{P4_1}
\begin{aligned}
	&\left( \mathrm{P4'}\right):\underset{\hat{\mathbf{V}}\in\mathcal{M}_{1}}{\mathrm{max}}\; \bar{f}_{4}(\hat{\mathbf{V}})=\\&\sum_{k=1}^{K}\frac{\tilde{\zeta} _{k}\left|\left(\sum_{s=1}^{2}\mathbf{g}_{s,k}^{H}\mathbf{\Phi }_{s}\mathbf{\hat{H}}_{s}+\mathbf{g}_{2,k}^{H}\mathbf{\Phi }_{2}\mathbf{\Lambda}_{1,2}\mathbf{\Phi }_{1}\mathbf{\hat{H}}_{1}\right)\hat{\mathbf{V}}\mathbf{E}_{kk}\right|^{2}}{\sum_{j=1}^{K}\left|\left(\sum_{s=1}^{2}\mathbf{g}_{s,k}^{H}\mathbf{\Phi }_{s}\mathbf{\hat{H}}_{s}+\mathbf{g}_{2,k}^{H}\mathbf{\Phi }_{2}\mathbf{\Lambda}_{1,2}\mathbf{\Phi }_{1}\mathbf{\hat{H}}_{1}\right)\hat{\mathbf{V}}\mathbf{E}_{kj}\right|^{2}+\sigma _{k}^{2}},
\end{aligned}
\end{equation}}
 and
 \begin{equation}\label{P5_1}
\begin{aligned}
&\left( \mathrm{P5'}\right):\;\underset{\mathbf{u}_x\in \mathcal{M}_{2}}{\mathrm{max}} \;\bar{f}_{5}(\mathbf{u}_x)\\&=\sum_{k=1}^{K}\frac{\tilde{\zeta }_{k}\left|\left(\mathbf{u}_1^{H}\mathbf{A}+\mathbf{u}_2^{H}\mathbf{B}\right)\mathbf{v}_{k}\right|^{2}}{\sum_{j=1}^{K}\left|\left(\mathbf{u}_1^{H}\mathbf{A}+\mathbf{u}_2^{H}\mathbf{B}\right)\mathbf{v}_{j}\right|^{2}+\sigma _{k}^{2}}
,
\end{aligned}
\end{equation}
where the definitions of $\mathbf{\hat{H}}_{1}$ and $\mathbf{\hat{H}}_{2}$ are similar to $\mathbf{\hat{H}}$. It should be noted that the forms of $\mathbf{A}$ and $\mathbf{B}$ are related to $x$ and are expressed as
\begin{equation}
    \mathbf{A} =
\begin{cases}
\mathrm{diag}(\mathbf{g}_{1,k}^{H}+\mathbf{g}_{2,k}^{H}\mathbf{\Phi }_{2}\mathbf{\Lambda}_{1,2})\mathbf{H}_1,  & x=1 \\
\mathrm{diag}(\mathbf{g}_{1,k}^{H})\mathbf{H}_1, & x=2
\end{cases}
\end{equation}
and
\begin{equation}
    \mathbf{B} =
\begin{cases}
\mathrm{diag}(\mathbf{g}_{2,k}^{H})\mathbf{H}_2,  & x=1 \\
\mathrm{diag}(\mathbf{g}_{2,k}^{H})(\mathbf{H}_2+\mathbf{\Lambda}_{1,2}\mathbf{\Phi }_{1}\mathbf{H}_1), & x=2.
\end{cases}
\end{equation}

Define $ \hat{\tilde{\mathbf{h}}}_{k}^{H}=\sum_{s=1}^{2}\mathbf{g}_{s,k}^{H}\mathbf{\Phi }_{s}\mathbf{\hat{H}}_{s}+\mathbf{g}_{2,k}^{H}\mathbf{\Phi }_{2}\mathbf{\Lambda}_{1,2}\mathbf{\Phi }_{1}\mathbf{\hat{H}}_{1} $, and then the Euclidean gradient of $f_4(\hat{\mathbf{V}})$ in \eqref{EV} and $f_5(\mathbf{u})$ in \eqref{Eu} becomes \eqref{EVinterIRC} and \eqref{EuInterIRS}, respectively, as shown at the top of next page.
\begin{figure*}
\begin{equation}\label{EVinterIRC}
\mathrm{Egrad}\;\bar{f}_{4}(\hat{\mathbf{V}})=\frac{\partial \bar{f}_{4}(\hat{\mathbf{V}})}{\partial \hat{\mathbf{V}}}=
\sum_{k=1}^{K}\tilde{\zeta }_{k}\cdot \left( \frac{2\hat{\tilde{\mathbf{h}}}_{k}^{H}\hat{\mathbf{V}}\mathbf{E}_{kk}\hat{\tilde{\mathbf{h}}}_{k}\mathbf{E}_{kk}^{H}}{\sum_{j=1}^{K}|\hat{\tilde{\mathbf{h}}}_{k}^{H}\hat{\mathbf{V}}\mathbf{E}_{kj}|^{2}+\sigma _{k}^{2}}
-2\sum_{i=1}^{K}\frac{|\hat{\tilde{\mathbf{h}}}_{k}^{H}\hat{\mathbf{V}}\mathbf{E}_{kk}|^{2}\hat{\tilde{\mathbf{h}}}_{k}^{H}\hat{\mathbf{V}}\mathbf{E}_{ki}\hat{\tilde{\mathbf{h}}}_{k}\mathbf{E}_{ki}^{H}}{(\sum_{j=1}^{K}|\hat{\tilde{\mathbf{h}}}_{k}^{H}\hat{\mathbf{V}}\mathbf{E}_{kj}|^{2}+\sigma _{k}^{2})^{2}}\right),
\end{equation}
\end{figure*}
\begin{figure*}
\hrule
\begin{equation}\label{EuInterIRS}
\begin{split}
&\mathrm{Egrad}\;\bar{f}_{5}(\mathbf{u }_1)=\\&\frac{\partial \bar{f}_{5}(\mathbf{u}_1)}{\partial \mathbf{u}_1}=\sum_{k=1}^{K}\tilde{\zeta }_{k}\cdot \left(\frac{2\left(\mathbf{v}_{k}^{H}\mathbf{A}^H\mathbf{u}_1\mathbf{A}\mathbf{v}_{k}+\mathbf{v}_{k}^{H}\mathbf{B}^H\mathbf{u}_2\mathbf{A}\mathbf{v}_{k}\right)}{\sum_{j=1}^{K}\left|\left(\mathbf{u}_1^{H}\mathbf{A}+\mathbf{u}_2^{H}\mathbf{B}\right)\mathbf{v}_{j}\right|^{2}+\sigma _{k}^{2}}-2\sum_{i=1}^{K}\frac{\left|\left(\mathbf{u}_1^{H}\mathbf{A}+\mathbf{u}_2^{H}\mathbf{B}\right)\mathbf{v}_{k}\right|^{2}\left(\mathbf{v}_{i}^{H}\mathbf{A}^H\mathbf{u}_1\mathbf{A}\mathbf{v}_{i}+\mathbf{v}_{i}^{H}\mathbf{B}^H\mathbf{u}_2\mathbf{A}\mathbf{v}_{i}\right)  }{\left(\sum_{j=1}^{K}\left|\left(\mathbf{u}_1^{H}\mathbf{A}+\mathbf{u}_2^{H}\mathbf{B}\right)\mathbf{v}_{j}\right|^{2}+\sigma _{k}^{2}\right)^{2}}\right),\\
&\mathrm{Egrad}\;\bar{f}_{5}(\mathbf{u }_2)=\\&\frac{\partial \bar{f}_{5}(\mathbf{u}_2)}{\partial \mathbf{u}_2}=\sum_{k=1}^{K}\tilde{\zeta }_{k}\cdot \left(\frac{2\left(\mathbf{v}_{k}^{H}\mathbf{B}^H\mathbf{u}_2\mathbf{B}\mathbf{v}_{k}+\mathbf{v}_{k}^{H}\mathbf{A}^H\mathbf{u}_1\mathbf{B}\mathbf{v}_{k}\right)}{\sum_{j=1}^{K}\left|\left(\mathbf{u}_1^{H}\mathbf{A}+\mathbf{u}_2^{H}\mathbf{B}\right)\mathbf{v}_{j}\right|^{2}+\sigma _{k}^{2}}-2\sum_{i=1}^{K}\frac{\left|\left(\mathbf{u}_1^{H}\mathbf{A}+\mathbf{u}_2^{H}\mathbf{B}\right)\mathbf{v}_{k}\right|^{2}\left(\mathbf{v}_{i}^{H}\mathbf{B}^H\mathbf{u}_2\mathbf{B}\mathbf{v}_{i}+\mathbf{v}_{i}^{H}\mathbf{A}^H\mathbf{u}_1\mathbf{B}\mathbf{v}_{i}\right)  }{\left(\sum_{j=1}^{K}\left|\left(\mathbf{u}_1^{H}\mathbf{A}+\mathbf{u}_2^{H}\mathbf{B}\right)\mathbf{v}_{j}\right|^{2}+\sigma _{k}^{2}\right)^{2}}\right),
\end{split}
\end{equation}
\hrule
\end{figure*}

Substituting \eqref{EVinterIRC} and \eqref{EuInterIRS} back to \textbf{Algorithm 1}, then replacing $\mathbf{u}^{(t)}$ by $\mathbf{u}_1^{(t)}$ and $\mathbf{u}_2^{(t)}$ in step 5 of \textbf{Algorithm 2}, we complete the optimization procedure.

\subsection{Adjust Non-ideal IRS}\label{E}
For non-ideal IRS, due to hardware limitations, it is impossible to set the reflection coefficient to an arbitrary value to meet the requirement of infinite resolution of the reflection angle. Therefore, in this subsection, the reflection angle is quantized  so that it can only be selected from the set of feasible angles with the shortest distance from the optimized value. Quantization is realized by
\begin{equation}
\begin{aligned}
&\bar{\phi }_{i}=\underset{\phi ^{'}\in \mathcal{F}}{\mathrm{argmin}}\;\left|\phi _{i}-\phi^{'}\right|,\\
&\mathcal{F}=\left\{0,\frac{2\pi }{Q},\dots \frac{2\pi (Q-1)}{Q}\right\},i = 1, \dots, SM,
\end{aligned}
\end{equation}
where $\bar{\phi }_{i} $, $\phi _{i} $ and $\phi^{'} $ are defined as the reflection angle of the final output, infinite resolution reflection angle obtained by \textbf{Algorithm 2} and a specific angle, respectively. $ \mathcal{F} $ represents the set of feasible angles of finite resolution controlled by a $ Q $-level quantizer as mentioned in Section \ref{System Model}.

\subsection{Convergence and Complexity}\label{Convergence1}
Given the limited transmission power $ P $, the weighted sum-rate has an upper-bound. Moreover, the steps of updating $ \bm{\zeta} $, $\mathbf{\hat{V}} $ and $\mathbf{u} $ are monotonically  non-decreasing. Thus the alternating procedure will converge.

Owing to the effect of quantization, updating reflection coefficients loses monotonically increasing property. Fortunately, the steps for updating  $ \bm{\zeta} $ and $\mathbf{V} $ are both monotonous, so the algorithm proposed in this paper still converges, which will be verified in Section \ref{Simulation Results}.

The overall complexity of the DOMALO algorithm is mainly determined by operations of  retraction, vector transport and Riemann gradient of $\hat{\mathbf{V}}$ and $\mathbf{u}$. Thus, we only consider these three steps in \textbf{Algorithm 1}. For $\hat{\mathbf{V}}$, the computational complexities of retraction, vector transport, and Riemann gradient, as shown in (\ref{RV}), (\ref{TV}) and (\ref{EV}), are $ \mathcal{O}\left( NK\right)  $, $ \mathcal{O}\left( NK^2\right)  $ and $ \mathcal{O}\left( KS^2M^2+KSMN+NK^3\right)  $, respectively. The computational complexities of $\mathbf{u}$, which come from the steps in (\ref{Ru}), (\ref{Tu}) and (\ref{Eu}), are $ \mathcal{O}\left( S^3M^3\right)  $, $ \mathcal{O}\left( S^2M^2\right)  $ and $ \mathcal{O}\left( S^2M^2NK+SMNK^2\right)  $. According to \cite{al},  the strict convergence of \textbf{Algorithm 1} remains an open problem. Hence, it is rather
intractable to analyze the iteration number needed for \textbf{Algorithm 1}. Note that \textbf{Algorithm 1} iterates $T_1$ and $T_2$ times before terminating for $\hat{\mathbf{V}}$ and $\mathbf{u}$, respectively.  Therefore, for \textbf{Algorithm 2}, the complexities of updating $\hat{\mathbf{V}}$ and $\mathbf{u}$ are $\mathcal{O}\left(T_1KSM\cdot\max{\{SM,N\}}+T_1NK^3\right)$ and $\mathcal{O}\left(T_2SM\cdot\max{\{SMNK,NK^2,S^2M^2\}}\right)$ per iteration, respectively.

%\textcolor{red}{$L$ is the number of NOLS paths, $i$ or $m$? please check. }

\section{Maximize Minimal-Rate}\label{Maximize Minimal-Rate}
In fact, maximizing the weighted sum-rate may lead to some users taking most of the system resources, while others experience unbearably low level of service. Therefore, we may choose to maximize the minimum received rate among all users in many practical scenarios for fairness.

\subsection{Problem Formulation}

Similar to the weighted sum-rate problem in Section \ref{PF}, we formulate the weighted minimal-rate optimization problem as
\begin{equation}
\begin{aligned}
\left( \mathrm{Q1}\right) :&\;\max_{\mathbf{V},\mathbf{\Phi} }\:\min_k\; g_{1,k}(\mathbf{V},\mathbf{\Phi} )=\omega _{k}\mathcal{R}_{k}
\\&\mathrm{s.t.}\; \mathrm{tr}(\mathbf{VV}^{H})\leq P
\\&\phi_{i}\in \mathcal{F},i=1,2,\dots,SM.
\end{aligned}
\end{equation}

We can find from (\ref{Rk}) that $ \mathcal{R}_{k} $ and $ r_{k} $ are positively correlated, so maximizing the minimal $ \mathcal{R}_{k} $ can be simplified to maximizing minimal $ r_{k} $ for all users. The original objective function can be replaced by $ g_{2,k}\left( \mathbf{V},\mathbf{\Phi} \right) =\omega _{k}r_{k} $. $ (\mathrm{Q1}) $ also has both the BS power constraint and the IRS reflection coefficient constraint.

\subsection{Reconstruction of Original Problem}

To solve $ (\mathrm{Q1}) $, we consider the following parametric problem
\begin{equation}
\begin{aligned}
\left( \mathrm{Q2}\right) :\;
&G\left( \tau\right) =\max_{\mathbf{V},\mathbf{\Phi} }\:
g_3\left(\mathbf{V},\mathbf{\Phi} \right)
%\min_k\; \omega_{k}\left|\mathbf{g} _{k}^{H} \mathbf{\Phi}  \mathbf{H} \mathbf{v} _{k}\right|^{2}\\
%&-\tau\left (  \sum_{j\ne k}^{K} \left|\mathbf{g} _{k}^{H} \mathbf{\Phi}  \mathbf{H} \mathbf{v} _{j}\right|^{2}+\sigma _k^2 \right )
\\&\mathrm{s.t.}\; \mathrm{tr}(\mathbf{VV}^{H})\leq P
\\&\phi{i}\in \mathcal{F},i=1,2,\dots,SM,
\end{aligned}
\end{equation}
where the objective function
\begin{equation}
    g_3\left(\mathbf{V},\mathbf{\Phi} \right)\triangleq\min_k \,\omega_{k}\left|\mathbf{g} _{k}^{H} \mathbf{\Phi}  \mathbf{H} \mathbf{v} _{k}\right|^{2}-\tau\left (  \sum_{j\ne k}^{K} \left|\mathbf{g} _{k}^{H} \mathbf{\Phi}  \mathbf{H} \mathbf{v} _{j}\right|^{2}+\sigma _k^2 \right )
    \label{eq_g3}
\end{equation}
represents the received power shortage or redundancy of the $ k $-th user to reach a weighted SINR of $ \tau $.

$ G\left( \tau\right) $ is continuous and strictly decreasing over  $[0,\infty)$, and the optimal value $ \tau^* $ of $ \left( \mathrm{Q2}\right) $ is finite and satisfies $ G\left( \tau^*\right)=0 $. Moreover, if $ G\left( \tau\right)=0, $ then $ \left( \mathrm{Q2}\right) $ and $ \left( \mathrm{Q1}\right) $ have the same set of optimal solution \cite{barros1996new}. Based on the Dinkelbach-type algorithm proposed in \cite{crouzeix1985algorithm}, we can create a non-decreasing
sequence $ \tau_t, t\geq1$, obtained through $t$ iterations,
%\textcolor{red}{($n$ means the index of transmit antenna)}
converging from above to the optimal value $ \tau^* $.
%Instead of solving original problem $ (\mathrm{Q1}) $, we have reconstructed $ (\mathrm{Q1}) $ and compute $ G(\tau) $ to find the optimal solution.
In addition, Since the objective function $ g_3\left(\mathbf{V},\mathbf{\Phi} \right)$ %=\min_k \,\omega_{k}\left|\mathbf{g} _{k}^{H} \mathbf{\Phi}  \mathbf{H} \mathbf{v} _{k}\right|^{2}-\tau\left (  \sum_{j\ne k}^{K} \left|\mathbf{g} _{k}^{H} \mathbf{\Phi}  \mathbf{H} \mathbf{v} _{j}\right|^{2}+\sigma _k^2 \right ) $
is non-differentiable and non-smooth to $ \mathbf{V} $ and $ \mathbf{\Phi} $, we substitute it with a smooth exponential penalty function $ \tilde{g}_3\left(\mathbf{V},\mathbf{\Phi} \right) $, yielding the following optimization problem \cite{xu2001smoothing}
\begin{equation}
\begin{aligned}
\left( \mathrm{Q3}\right) :\;
&\max_{\mathbf{V},\mathbf{\Phi} }\;\tilde{g}_3\left(\mathbf{V},\mathbf{\Phi} \right)
\\&\mathrm{s.t.}\; \mathrm{tr}(\mathbf{VV}^{H})\leq P
\\&\phi{i}\in \mathcal{F},i=1,2,\dots,SM,
\end{aligned}
\end{equation}
where
\begin{equation}
\begin{aligned}
\tilde{g}_3\left(\mathbf{V},\mathbf{\Phi}\right) \triangleq
&-\mu\log\Biggl(\sum_{k=1}^{K}\exp\Biggl(-\omega_{k}\left|\mathbf{g} _{k}^{H} \mathbf{\Phi}  \mathbf{H} \mathbf{v} _{k}\right|^{2}/\mu\\
&+\tau\left (  \sum_{j\ne k}^{K} \left|\mathbf{g} _{k}^{H} \mathbf{\Phi}  \mathbf{H} \mathbf{v} _{j}\right|^{2}+\sigma _k^2 \right )/\mu\Biggr)\Biggr),
\end{aligned}
\end{equation}
and $ \mu>0 $ is a smoothing parameter satisfying
\begin{equation}
\tilde{g}_3\left(\mathbf{V},\mathbf{\Phi}\right) \leq g_3\left(\mathbf{V},\mathbf{\Phi} \right)\leq\tilde{g}_3\left(\mathbf{V},\mathbf{\Phi}\right)+\mu\log K.
\end{equation}
%where $K$ is the number of users. (already defined)

As $ \mu $ decreases, the gap between $ \tilde{g}_3\left(\mathbf{V},\mathbf{\Phi}\right) $ and $ g_3\left(\mathbf{V},\mathbf{\Phi}\right)  $ decreases. Then, the  problem $ \left( \mathrm{Q3}\right) $ can well approximate the problem $ \left( \mathrm{Q2}\right) $ when $ \mu $ is set appropriately. However, when the approximation accuracy is high, the smooth approximating problem becomes significantly ill-conditioned. Consequently, using smooth exponential penalty function is complicated due to  the need for trading-off accuracy of approximation against problem ill-conditioning \cite{polak2003algorithms}.

To solve $ (\mathrm{Q3}) $, we alternately optimize $ \mathbf{V} $ and $ \mathbf{\Phi} $, while carefully adjusting the value of $ \mu $. %Details will be explained below.

\subsection{Fix $\mathbf{\Phi} $ and Optimize $ \mathbf{V} $}\label{2C}
For any given reflection matrix $\mathbf{\Phi} $, the beamforming matrix $\mathbf{V} $ in $ \left( \mathrm{Q3}\right) $  can be optimized by solving the following problem
\begin{equation}
\begin{aligned}
\left( \mathrm{Q3'}\right) :\;
&\max_{\mathbf{V} }\;\tilde{g}_3\left(\mathbf{V} \right)\\
&=-\mu_\mathbf{{V}}\log\Biggl(\sum_{k=1}^{K}\exp\Biggl(-\omega_{k}\left|\mathbf{g} _{k}^{H} \mathbf{\Phi}  \mathbf{H} \mathbf{v} _{k}\right|^{2}/\mu_\mathbf{{V}}\\
&+\tau\left (  \sum_{j\ne k}^{K} \left|\mathbf{g} _{k}^{H} \mathbf{\Phi}  \mathbf{H} \mathbf{v} _{j}\right|^{2}+\sigma _k^2 \right )/\mu_\mathbf{{V}}\Biggr)\Biggr)
\\&\mathrm{s.t.}\; \mathrm{tr}(\mathbf{VV}^{H})\leq P,
\end{aligned}
\end{equation}
where $ \mu_\mathbf{{V}} $ is the smoothing parameter related to $ \mathbf{{V}} $.

Similar to Section \ref{C}, we define a manifold $ \mathcal{M}_{1}=\left\lbrace  \mathbf{\hat{V}}\in \mathbb{C}^{(N+1)\times K}\big| \mathrm{tr}(\mathbf{\hat{V}\hat{V}}^{H})= 1 \right\rbrace $. Then, this sub-problem can be solved using manifold optimization and rewritten as
\begin{equation}\label{Q4}
\begin{aligned}
\left( \mathrm{Q4}\right):&\;\max_{\mathbf{\hat{V}}\in\mathcal{M}_1}\;\tilde{g}_4\left(\mathbf{\hat{V}}\right)=-\mu_\mathbf{\hat{V}}\log\left(\sum_{k=1}^{K} \psi_k\left( \mathbf{\hat{V}}\right)\right)  \\=
&-\mu_\mathbf{\hat{V}}\log\Biggl(\sum_{k=1}^{K}\exp\Biggl(-\omega_{k}\left|\mathbf{g} _{k}^{H} \mathbf{\Phi}  \mathbf{\hat{H}} \mathbf{\hat{V}} \mathbf{E}_{kk}\right|^{2}/\mu_\mathbf{\hat{V}}\\
&+\tau\left (  \sum_{j\ne k}^{K} \left|\mathbf{g} _{k}^{H} \mathbf{\Phi}   \mathbf{\hat{H}} \mathbf{\hat{V}} \mathbf{E}_{kj}\right|^{2}+\sigma _k^2 \right )/\mu_\mathbf{\hat{V}}\Biggr)\Biggr),
\end{aligned}
\end{equation}
where $ \mathbf{\hat{H}}=\sqrt{P}\left[ \mathbf{H},\mathbf{0}\right]  $.
%\textcolor{red}{how $\hat{\textbf{H}}$ comes from?}
The tangent plane, retraction, and vector transport on $ \mathcal{M}_1 $ are the same as (\ref{PV}), (\ref{RV}) and (\ref{TV}), respectively. Introducing $ \tilde{\mathbf{h}}_{k}^{H}=\mathbf{g}_{k}^{H}\mathbf{\Phi }\hat{\mathbf{H}} $ mentioned in Section \ref{C}, the Riemann gradient can be calculated using (\ref{REV}) and (\ref{EV2}), shown at the top of next page.
\begin{figure*}[ht]
	\begin{equation}\label{EV2}
	\mathrm{Egrad}\;\tilde{g}_4(\hat{\mathbf{V}})=\frac{\partial \tilde{g}_{4}(\hat{\mathbf{V}})}{\partial \hat{\mathbf{V}}}	=\frac{\sum_{k=1}^{K}\psi_k\left( \mathbf{\hat{V}}\right)\left ( 2\omega_k\tilde{\mathbf{h}}_{k}^{H}\hat{\mathbf{V}}\mathbf{E}_{kk}\tilde{\mathbf{h}}_{k}\mathbf{E}_{kk}^{H} -2\tau\sum_{j\neq k}^{K}\tilde{\mathbf{h}}_{k}^{H}\hat{\mathbf{V}}\mathbf{E}_{kj}\tilde{\mathbf{h}}_{k}\mathbf{E}_{kj}^{H}\right )}{\sum_{k=1}^{K}\psi_k\left( \mathbf{\hat{V}}\right)}.
	\end{equation}
	\hrule
\end{figure*}

According to the formula described above, we can apply \textbf{Algorithm 1} to solve this subproblem.

At the $ t $-th iteration, when we obtain the optimal $ \mathbf{V}^{(t)} $, to increase weighted minimal-rate monotonically, we need to decide whether or not to update $ \tau $. Although we can use \textbf{Algorithm 1} to ensure that $ \tilde{g}_4\left(\mathbf{\hat{V}}^{(t)}\right)\geq\tilde{g}_4\left(\mathbf{\hat{V}}^{(t-1)}\right) $ is satisfied, we cannot guarantee $ {g}_3\left(\mathbf{{V}}^{(t)}\right)\geq {g}_3\left(\mathbf{{V}}^{(t-1)}\right) $. In fact, the iterative results of $ {g}_3\left(\mathbf{{V}}\right)  $ and the iterative results of $ \tilde{g}_4\left(\mathbf{\hat{V}}\right) $ are divergent because the corresponding $ \mu_\mathbf{\hat{V}} $ is relatively large, and $ \tilde{g}_4\left(\mathbf{\hat{V}}\right) $ doesn't fully approximate $ {g}_3\left(\mathbf{{V}}\right)  $. For this reason, we use (\ref{t}) to update $ \tau $ only when $ {g}_3\left(\mathbf{{V}}^{(t)}\right)\geq {g}_3\left(\mathbf{{V}}^{(t-1)}\right) $ holds, that is
\begin{equation}\label{t}
\tau=\min_k \,\frac{\omega_{k}\left|\mathbf{g} _{k}^{H} \mathbf{\Phi}  \mathbf{H} \mathbf{v} _{k}\right|^{2}}{\left (  \sum_{j\ne k}^{K} \left|\mathbf{g} _{k}^{H} \mathbf{\Phi}  \mathbf{H} \mathbf{v} _{j}\right|^{2}+\sigma _k^2 \right )}.
\end{equation}

Otherwise, if $ {g}_3\left(\mathbf{{V}}^{(t)}\right)< {g}_3\left(\mathbf{{V}}^{(t-1)}\right) $,
 %\textcolor{red}{(I can't understand otherwise to what? the uptade of $t$?)}
 we reduce $ \mu_\mathbf{\hat{V}} $ by $ \mu_\mathbf{\hat{V}} =\varsigma\mu_\mathbf{\hat{V}} $ where $ \varsigma\in(0,1) $. Note that $ \tau $ is the optimal value for the objective function of the optimization problem $\left(  \mathrm{Q4}\right)  $ and denoted as $ \tau_{beam}^{(t)} $ at the $t$-th iteration. The relevant steps are detailed in \textbf{Algorithm 3}.

\subsection{Fix $\mathbf{V} $ and Optimize $ \mathbf{\Phi} $}\label{2D}

%In this subsection, we optimize $ \mathbf{\Phi} $ by fixing $\mathbf{V} $.
Similar to the steps in Section \ref{D}, we define vector $\mathbf{u} $ that satisfies $\mathbf{u}^{*}=\mathrm{vec}(\mathbf{\Phi}) $. Then, we have $\mathbf{u}^{H}\mathrm{diag}(\mathbf{g}_{k}^{H})\mathbf{H}=\mathbf{g}_{k}^{H}\mathbf{\Phi }\mathbf{H} $. Through forming a manifold $ \mathcal{M}_{2}=\left\lbrace  \mathbf{u}\in \mathbb{C}^{M\times 1}\big|(\mathbf{u}\mathbf{u}^H)_{ii} = 1, i = 1:SM   \right\rbrace $, $ \mathbf{\Phi} $ in $ \left( \mathrm{Q3}  \right) $ can be optimized by solving the following problem
\begin{equation}\label{Q5}
\begin{aligned}
\left( \mathrm{Q5}\right) :\;
&\max_{\mathbf{u}\in\mathcal{M }_2}\;\tilde{g}_5\left(\mathbf{u} \right)=-\mu_\mathbf{u}\log\left(\sum_{k=1}^{K} \rho_k\left( \mathbf{u}\right)\right) \\
&=-\mu_\mathbf{u}\log\Biggl(\sum_{k=1}^{K}\exp\Biggl(-\omega_{k}\left|\mathbf{u}^{H}\mathrm{diag}(\mathbf{g}_{k}^{H})\mathbf{H} \mathbf{v} _{k}\right|^{2}/\mu_\mathbf{u}\\
&+\tau\left (  \sum_{j\ne k}^{K} \left|\mathbf{u}^{H}\mathrm{diag}(\mathbf{g}_{k}^{H})\mathbf{H} \mathbf{v} _{j}\right|^{2}+\sigma _k^2 \right )/\mu_\mathbf{u}\Biggr)\Biggr),
\end{aligned}
\end{equation}
where $ \mu_\mathbf{u} $ is the smoothing parameter related to $ \mathbf{u} $.

The tangent plane, retraction and vector transport on $ \mathcal{M}_2 $ are defined in (\ref{Pu}), (\ref{Ru}) and (\ref{Tu}), respectively. The Riemann gradient can be calculated using (\ref{REu}) and (\ref{Eu2}), shown at the top of next page.
\begin{figure*}
	\begin{equation}\label{Eu2}
	\begin{split}
	&\mathrm{Egrad}\;\tilde{g}_5(\mathbf{u})=\frac{\partial \tilde{g}_{5}(\mathbf{u})}{\partial \mathbf{u}}	=\\
	&\frac{1}{\sum_{k=1}^{K}\rho_k\left( \mathbf{u}\right)}\cdot\sum_{k=1}^{K}\rho_k\left( \mathbf{u}\right)\left ( 2\omega_k\mathbf{v}_{k}^{H}\mathbf{H}^{H}\mathrm{diag}^{H}(\mathbf{g}_{k}^{H})\mathbf{u}\mathrm{diag}(\mathbf{g}_{k}^{H})\mathbf{H}\mathbf{v}_{k}-2\tau\sum_{j\neq k}^{K} \mathbf{v}_{j}^{H}\mathbf{H}^{H}\mathrm{diag}^{H}(\mathbf{g}_{k}^{H})\mathbf{u}\mathrm{diag}(\mathbf{g}_{k}^{H})\mathbf{H}\mathbf{v}_{j}\right ).
	\end{split}
	\end{equation}
	\hrule
\end{figure*}

Similarly, we can apply \textbf{Algorithm 1} to solve this sub-problem. The update conditions for $ \tau $ and $ \mu_\mathbf{u} $ are similar to those given in Section \ref{2C}.  $ \tau $ is also the optimal value for the objective function of the problem $\left(  \mathrm{Q5}\right)  $ and is denoted as $ \tau_{phase}^{(t)} $ at the $t$-th iteration. %The relevant steps are explained in \textbf{Algorithm 3}.
\subsection{Alternating Optimization Framework}

Based on the two manifold optimization approaches introduced in Sections \ref{2C} and \ref{2D}, we then consider the complete alternating optimization algorithm optimizing the
beamforming vector and the reflection matrix, called the smooth double manifold alternating optimization (S-DOMALO), as summarized in \textbf{Algorithm 3}. As mentioned above, the optimization variables are divided into two blocks, i.e., $ \left\lbrace \mathbf{v_k}\right\rbrace  $ and  $ \mathbf{\Phi} $, where $ \left\lbrace \mathbf{v_k}\right\rbrace  $ are lifted to $ \mathbf{\hat{V}} $ and $ \mathbf{\Phi} $ is vectorized to $ \mathbf{u} $. Then, $ \mathbf{\hat{V}} $ and $ \mathbf{u} $ are optimized by fixing the other one alternately.

\begin{algorithm}[htbp]
	\caption{Smooth DOuble Manifold ALternating Optimization Algorithm}
	\begin{algorithmic}[1]
		\REQUIRE Channel state information about $ \mathbf{H} $ and $ \left\lbrace \mathbf{g}_{k}\right\rbrace $; user service priority  $ \omega _{k},\forall k $.
		
		\STATE Initialize $ \mathbf{\hat{V}}^{(0)}$ and  $ \mathbf{u}^{(0)} $;
		\STATE Calculate $ \tau^{(0)} $ by (\ref{t});
		\FOR {$t=1,2,...$}
		\STATE Update $ \mathbf{\hat{V}}^{(t-1)}$ to solve $(\mathrm{Q4})$ using \textbf{Algorithm 1};
		\IF {$ {g}_3\left(\mathbf{{V}}^{(t)}\right)\geq {g}_3\left(\mathbf{{V}}^{(t-1)}\right) $}
		\STATE Update $ \tau^{(t)} $ by (\ref{t});
		\ELSE
		\STATE $ \mathbf{\hat{V}}^{(t)}=\mathbf{\hat{V}}^{(t-1)}$, $ \tau^{(t)}=\tau^{(t-1)} $, $ \mu_\mathbf{\hat{V}}=\varsigma\mu_\mathbf{\hat{V}} $;
		\ENDIF
		\STATE Update $ \mathbf{u}^{(t-1)}$ to solve $(\mathrm{Q5})$ using \textbf{Algorithm 1};
		\IF {$ {g}_3\left(\mathbf{{u}}^{(t)}\right)\geq {g}_3\left(\mathbf{{u}}^{(t-1)}\right) $}
		\STATE Update $ \tau^{(t)} $ by (\ref{t});
		\ELSE
		\STATE $ \mathbf{u}^{(t)}=\mathbf{u}^{(t-1)}$,  $ \tau^{(t)}=\tau^{(t-1)} $, $ \mu_\mathbf{u}=\varsigma\mu_\mathbf{u} $;
		\ENDIF
		\IF {$ \mu_\mathbf{\hat{V}}\leq\varepsilon_0  $ or $ \mu_\mathbf{u}\leq\varepsilon_0  $ or $t>T_{max}$ }
		\BREAK
		\ENDIF
		\STATE $ t=t+1 $;
		\ENDFOR
		\ENSURE Optimized beamforming vecter $ \left\lbrace \mathbf{v}_{k}\right\rbrace $ and reflection matrix $ \mathbf{\mathbf{\Phi}} $, optimal value of $ g_{3}(\mathbf{V}^{(t)},\mathbf{\Phi}^{(t)} ) $.
	\end{algorithmic}
\end{algorithm}

\textbf{Remark 1}: To have more degrees of freedom, we divide the smoothing parameter $ \mu $ into $ \mu_{\mathbf{\hat{V}}} $ and $ \mu_{\mathbf{u}} $.

\textbf{Remark 2}: In practice, the initial values of $ \mu_{\mathbf{\hat{V}}} $ and $ \mu_{\mathbf{u}} $
cannot be too large to prevent $ \tilde{g}_4\left(\mathbf{\hat{V}}\right)$ and $\tilde{g}_5\left(\mathbf{u} \right) $ from losing information of the objective $ g_{3}(\mathbf{V},\mathbf{\Phi} ) $. In extreme cases, $ \tilde{g}_4\left(\mathbf{\hat{V}}\right)\approx-\mu_\mathbf{\hat{V}}\log K$ and $\tilde{g}_5\left(\mathbf{u} \right)\approx-\mu_\mathbf{u}\log K $ are satisfied if $ \mu_{\mathbf{\hat{V}}} $ and $ \mu_{\mathbf{u}} $ are very large.

\textbf{Remark 3}: Steps 16-18 in \textbf{Algorithm 3} ensure that $ \mu_{\mathbf{\hat{V}}} $ and $ \mu_{\mathbf{u}} $ are not too small. If the smoothing parameters are too small, the smooth approximating problem becomes significantly ill-conditioned, and \textbf{Algorithm 1} cannot run successfully. For any $ \mathbf{\hat{V}} $ and $ \mathbf{u} $, the output value of $ \tilde{g}_4\left(\mathbf{\hat{V}}\right)$ and $\tilde{g}_5\left(\mathbf{u} \right) $ are always $+\infty$.

\textbf{Remark 4}: The situation of considering inter-IRS channel case is the same as in \ref{InterIRS}.

\subsection{Convergence and Complexity}
At the arbitrary $ t $-th iteration, since $  \mathbf{\hat{V}}^{(t)}  $  and $\mathbf{u}^{(t)}  $ are all optimized solutions, we have $ \tau_{phase}^{(t)}\geq \tau_{beam}^{(t)}\geq \tau_{phase}^{(t-1)} $.
Hence, we can obtain a monotonically non-decreasing sequence of $ \tau $. Due to the limited transmission power $ P $, the weighted minimal-rate is upper-bounded, and thus the alternating procedure will converge.
Although the difference between \textbf{Algorithm 2} and \textbf{Algorithm 3} is relatively significant, their computational complexity is almost the same. Specifically, both of them mainly depend on operations of  retraction, vector transport and Riemann gradient of $\hat{\mathbf{V}}$ and $\mathbf{u}$. For one iteration of \textbf{Algorithm 3}, the complexities are $\mathcal{O}\left(T_1KSM\cdot\max{\{SM,N\}}+T_1NK^3\right)$ and $\mathcal{O}\left(T_2SM\cdot\max{\{SMNK,NK^2,S^2M^2\}}\right)$ when \textbf{Algorithm 1} iterates $T_1$ and $T_2$ times to terminate for $\hat{\mathbf{V}}$ and $\mathbf{u}$, respectively.

\section{Simulation Results}\label{Simulation Results}

%In this section, we provided many simulation results under different conditions to evaluate the performance of the proposed algorithms.

Simulations are performed in this section to demonstrate the performances of the proposed algorithms. For BS and all IRSs, the number of antennas in each row is fixed to $5$, we can adjust only the number of column antennas. The variance of AWGN for each user is set as $\sigma^{2}=\sigma^{2}_k=-80\mathrm{dBm}, k=1,\cdots,K $.  For simplicity, all of the weights $ \omega_k $ are set to $1$.

By employing the classic Saleh-Valenzula channel model \cite{el2014spatially}, we have
%$ \mathbf{H}_{s} $ and $ \mathbf{g}_{s,k}^{H} $ can be expressed as:
\begin{equation}
	\mathbf{H}_{s}=\sqrt{\frac{N M}{\varrho_\mathrm{BI,s} }} \sum_{l=0}^{L_{BI,s}} \beta_{l} \mathbf{a}_{\mathrm{IRS}}\left(\varphi_{s,l}^{r}, \theta_{s,l}^{r}\right) \mathbf{a}_{\mathrm{BS}}\left(\varphi_{ l}^{t}, \theta_{l}^{t}\right)^{H},
	\label{h}
\end{equation}

\begin{equation}
	\mathbf{g}_{s,k}^{H}=\sqrt{\frac{ M}{\varrho_\mathrm{IU,s} }} \sum_{l=0}^{L_{IU,s}} \beta_{l} \mathbf{a}_{\mathrm{IRS}}\left(\varphi_{s, l}^{t}, \theta_{s,l}^{t}\right)^{H}, \label{gh}
\end{equation}
and
\begin{equation}
	\mathbf{\Lambda}_{1,2}=\sqrt{\frac{ M^2}{\varrho_\mathrm{I,1,2} }} \sum_{l=0}^{L_{I,1,2}} \beta_{l} \mathbf{a}_{\mathrm{IRS}}\left(\varphi_{2,l}^{r}, \theta_{2,l}^{r}\right) \mathbf{a}_{\mathrm{IRS}}\left(\varphi_{ 1,l}^{t}, \theta_{1,l}^{t}\right)^{H},
	\label{Lambda}
\end{equation}
where $ l = 0 $ represents the line-of-sight (LoS) path, $ L_{BI,s} $, $ L_{IU,s} $ and $ L_{I,1,2} $ are the number of NLoS paths; $ \varrho_\mathrm{BI,s} $, $ \varrho_\mathrm{IU,s} $ and $\varrho_\mathrm{I,1,2}$ denote the path-loss between BS-IRS, IRS-users and IRS1-IRS2, respectively;
$ \beta_{l} $ is the complex gain of $ l $-th path. %\textcolor{red}{(the same as the auxiliary variable in section III-B, similarly, $\theta$ the same meaning?)}
Here, the azimuth and elevation angles at receiver and transmitter are denoted by $ \varphi_{ l}^{r}, \theta_{l}^{r} $ and $ \varphi_{ l}^{t}, \theta_{l}^{t} $, respectively; %\textcolor{red}{(\textit{To avoid confusion, it is worth mentioning that $ \alpha $ and $ \theta $ only mean the complex gain and the elevation angle in \ref{System Model}. In the following sections, they have other meanings.})you have to change the symbol. it's not allowed the double meaning of the same variable, it's suggested that you change $\phi$ to be $\varphi$, change $\alpha$ to $\beta$ here}
%In \eqref{h} and \eqref{gh},
$ \mathbf{a}_{\mathrm{IRS}}\left(\varphi, \theta\right)$ and $ \mathbf{a}_{\mathrm{BS}}\left(\varphi, \theta\right) $ are the steering vectors of IRS and BS. Assuming the antenna elements are arranged as a uniform planar array (UPA) spaced by $d=\lambda/2$, we have
\begin{equation}
	\begin{aligned}
	&\mathbf{a}(\varphi, \theta)= \frac{1}{\sqrt{R C}}\\
	&\times\left [e^{j\frac{2 \pi}{\lambda} d\bigl(0\sin (\varphi) \sin (\theta)\bigr)},\cdots, e^{j\frac{2 \pi}{\lambda} d\bigl((R-1)\sin (\varphi) \sin (\theta)\bigr)} \right ]^{T}\\
	& \otimes\left [ e^{j\frac{2 \pi}{\lambda} d\bigl(0\cos (\theta)\bigr)},\cdots, e^{j\frac{2 \pi}{\lambda} d\bigl((C-1)\cos (\theta)\bigr)} \right ]^{T}	,
	\end{aligned}
\end{equation}
where $ R $ and $ C $ represent the number of antennas in each row and column, respectively.
%$ \lambda $ is the wavelength and $ d=\lambda/2 $ is the distance between adjacent antenna elements.

For $ \mathbf{H}_s $ and $ \mathbf{g}^H_{s,k} $, $ s=(1,\cdots,S) $ and $ k=(1,\cdots,K) $, there are a total of $ S(K+1) $ channels, with $  L=3 $ scattering paths and $ \beta_l\sim\mathcal{CN} (0,0.4) $. In addition, $ \beta_0 $ satisfies $ \beta_0\sim\mathcal{CN} (0,2) $. All azimuth and elevation angles are uniformly distributed in $ [0,2\pi) $ and $ [0,\pi) $, respectively. The path-loss including $ \varrho_\mathrm{I,1,2} $, $ \varrho_\mathrm{BI,s} $ and $ \varrho_\mathrm{IU,s} $ is given by
\begin{equation}
\varrho =\left(\frac{4\pi f_{\mathrm{carrier}}D}{c}\right)^{2},
\end{equation}
where $ c $ is the speed of light, $ D $ %\textcolor{red}{$d=\lambda/2$is defined before as the space of antennas}
is the distance between any two points, and $ f_{\mathrm{carrier}} $ is the carrier frequency and set to 3GHz.
%In addition, these settings also apply to $\mathbf{\Lambda}_{1,2}$.
The coordinate of the base station is $(0, 0)$, and those of IRS1 and IRS2 are $(10,24)$ and $(24,10)$, respectively. Users are uniformly distributed in a circular area with a radius of $2$ meters and a center location at $(20, 0)$.
$500$ Monte Carlo simulations are performed with channel realizations under different conditions.

The proposed algorithms %applied to weighted sum-rate problem and weighted minimal-rate problem
are compared with the following six benchmark schemes.
\begin{itemize}
\item \textbf{Scheme with random reflection matrix}: The reflection matrix is randomly generated, while $\mathbf{V}$ is updated by solving problem $ \left( \mathrm{P4}\right)  $ or $ \left( \mathrm{Q4}\right)  $.

\item \textbf{Alternating optimization with MRT}:
The beamforming matrix is set according to the MRT principle, that is
\begin{equation}\label{MRT}
\mathbf{V}=\sqrt{\frac{P}{\left\| \mathbf{H}_{MRT} \right\| ^2_F } }\mathbf{H}_{MRT}^H,
\end{equation}
where $ \mathbf{H}_{MRT}=\mathbf{g}_{k}^{H}\mathbf{\Phi }\mathbf{H} $, while the reflection matrix is updated by solving problem $ \left( \mathrm{P5}\right)  $ or $ \left( \mathrm{Q5}\right)  $. We optimize the  beamforming matrix and the reflection matrix in an alternating manner.

\item \textbf{Alternating optimization with ZF}: The beamforming matrix is set according to the ZF principle, that is
\begin{equation}\label{ZF}
\mathbf{V}=\sqrt{\frac{P}{\mathrm{tr}\left(\left( \mathbf{H}_{ZF}\mathbf{H}_{ZF}^H \right) ^{-1}\right)  } }\mathbf{H}_{ZF}^H\left( \mathbf{H}_{ZF}\mathbf{H}_{ZF}^H \right) ^{-1},
\end{equation}
where $ \mathbf{H}_{ZF}=\mathbf{g}_{k}^{H}\mathbf{\Phi }\mathbf{H} $. The reflection matrix is obtained by solving problem $ \left( \mathrm{P5}\right)  $ or $ \left( \mathrm{Q5}\right)  $. The  beamforming matrix and the reflection matrix are optimized alternately.
\item \textbf{Alternating optimization with MMSE}\cite{cho2010mimo}: The beamforming matrix is optimized based on MMSE, that is
\begin{equation}\label{MMSE}
\begin{split}
	\mathbf{V}&=\sqrt{\frac{P}{\mathrm{tr}\left(\mathbf{F}\mathbf{F}^H\right)  } }\mathbf{F},\\
	\mathbf{F}&=\mathbf{H}_{MMSE}^H\left( \mathbf{H}_{MMSE}\mathbf{H}_{MMSE}^H+\frac{\sigma^2K}{P}\mathbf{I}_K \right) ^{-1}_,
\end{split}
\end{equation}
where $ \mathbf{H}_{MMSE}=\mathbf{g}_{k}^{H}\mathbf{\Phi }\mathbf{H} $. The reflection matrix is obtained by solving problem $ \left( \mathrm{P5}\right)  $ or $ \left( \mathrm{Q5}\right)  $. The overall optimization framework is the same as the benchmarks using MRT or ZF.
\item \textbf{AIF algorithm }\cite{cao2020intelligent}:  The algorithm is proposed for the weighted sum-rate maximization problem. The beamforming matrix and the reflection matrix are optimized alternately with closed-form expressions, and related auxiliary variables are obtained using the Lagrangian multiplier method, bisection search method,  Schur complement and CVX toolbox \cite{cvx}.

\item \textbf{SOCP-SDR algorithm}\cite{9348016}: The algorithm is proposed for the weighted minimal-rate maximization problem and is based on alternating optimization. The authors optimize beamforming vectors via the second-order cone problem (SOCP) and reflection vector by using SDR.

\end{itemize}

In the initialization step of \textbf{Algorithm 2} and \textbf{Algorithm 3}, $ \mathbf{u}^{(0)} $ is set to $\mathbf{1}_M$. $\mathbf{\hat{V}}^{(0)}$ is calculated by (\ref{MRT}), but the only difference is that we use $\mathbf{\hat{H}}/\sqrt{P}$ instead of $\mathbf{H}_{MRT}$.

\subsection{Impact of the inter-IRS channels}

%Firstly, we utilize proposed DOMALO and S-DOMALO algorithms to explore the impact of the inter-IRS channels on system performance in this subsection,
To explore the impact of the inter-IRS channels on system performance, we first compare the performances of co-located multiple IRSs and separated multiple IRSs.
Similar to \cite{9362274}, we investigate the gain generated by double/multi-IRS.
Assuming that $M_1+M_2=90$ is always established, where $M_1$ and $M_2$ are the number of reflecting elements of IRS1 and IRS2, respectively.
%For comparison, two single-IRS cases are considered, in other words, $M_1=90$ and $M_2=90$.
The co-located multiple IRSs could be considered as single IRS cases with $M_1=90$ or $M_2=90$.
The number of BS antennas is fixed at $ N=20 $, the number of users is $ K=4 $, and the maximum transmission power is set as $ P=30\mathrm{dBm} $.

\begin{figure}[htbp]
	\centerline{\includegraphics[width=0.45\textwidth]{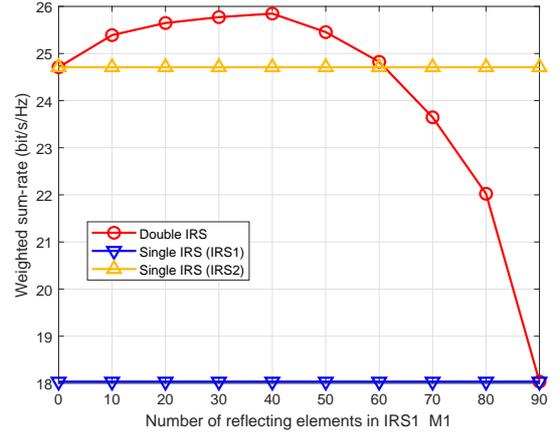}}
	\caption{Weighted sum-rate against the number of reflecting elements of IRS1.}
	\label{ds1}
\end{figure}
\begin{figure}[htbp]
	\centerline{\includegraphics[width=0.45\textwidth]{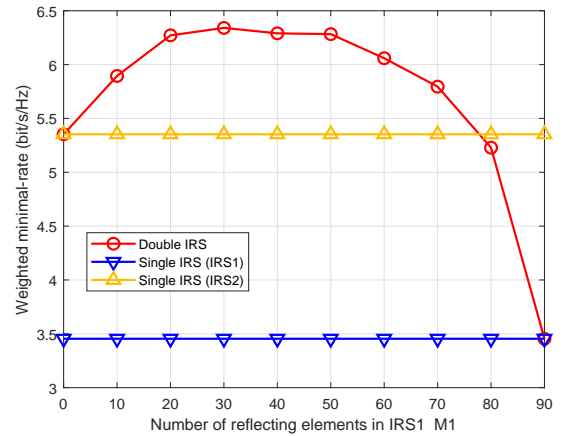}}
	\caption{Weighted minimal-rate against the number of reflecting elements of IRS1.
	}
	\label{ds2}
\end{figure}

As shown in Figs. \ref{ds1}-\ref{ds2}, the rate at which the red line exceeds the orange line is the additional gain obtained, which shows enormous advantage of using the multiple separated IRSs strategy over the single IRS strategy, even when the number of reflecting elements is constant.
Actually, the separated multiple IRSs strategy can obtain additional gain, since the spatial diversity and the non-correlation of channels can be utilized to provide cooperative beamforming gain.
%not only \textcolor{red}{because of the cooperation between IRSs (inter-IRS channels), but also the non-correlation of channels. (it's confusing)}

Considering the separated IRSs case, we then examine the impact of inter-IRS channels.
Let $p_i$ denote the probability of channel $h$ blocked every $10$ meters, that is

\begin{equation}
    p (h\text{~is~blocked}) =1-(1-p_i)^{\frac{d}{10}},
\end{equation}
where $d$ is the distance of inter IRS channel, and 
\begin{equation*}
    p_i= \left\{
    \begin{array}{cc}
        p_1, &\text{if}~h \in \{\mathbf{H}_1, \mathbf{\Lambda}_{1,2}\}, \\
        p_2, &\text{if}~h \in \{\mathbf{H}_2,  \mathbf{g}_{2,k}^H, \mathbf{g}_{1,k}^H \}.
    \end{array}
    \right.
\end{equation*}

For the convenience of subsequent comparison, the following three optimization and verification schemes are defined.
\begin{itemize}
    \item \textbf{Scheme 1}: In optimizing $\mathbf{V}$ and $\mathbf{\Phi}$, the inter-IRS channels are not considered, and in the performance verification stage, the inter-IRS channels are not considered either. That is to say, (\ref{eq_SINR}) is employed to construct the optimization problems during the optimization process and for calculating the weighted sum/minimal-rate during the verification stage.
    \item \textbf{Scheme 2}: In optimizing $\mathbf{V}$ and $\mathbf{\Phi}$, the inter-IRS channels are not considered, but in the performance verification stage, the inter-IRS channels are considered. Specifically, (\ref{eq_SINR}) is used to construct the optimization problems during the optimization process, but (\ref{eq_SINR2}) is used when calculating the weighted sum/minimal-rate and verifying the performance.
     \item \textbf{Scheme 3}: In optimizing $\mathbf{V}$ and $\mathbf{\Phi}$, the inter-IRS channels are considered, and in the performance verification stage, the inter-IRS channels are also considered. In other words,  we apply (\ref{eq_SINR2}) to construct the optimization problems and calculate the weighted sum/minimal-rate.
\end{itemize}

\begin{figure}[htbp]
	\centerline{\includegraphics[width=0.45\textwidth]{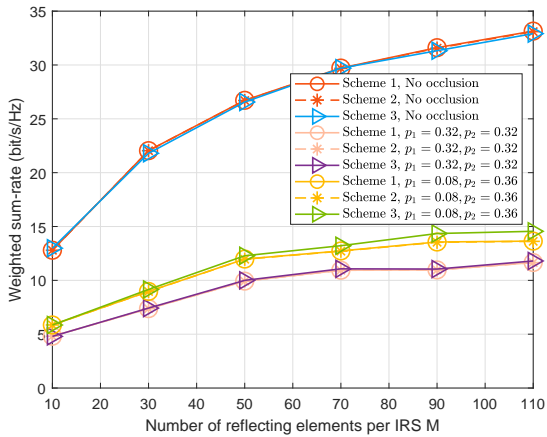}}
	\caption{Weighted sum-rate against the number of reflecting elements per IRS under the condition that the channels may be blocked.}
	\label{zhedang1}
\end{figure}
\begin{figure}[htbp]
	\centerline{\includegraphics[width=0.45\textwidth]{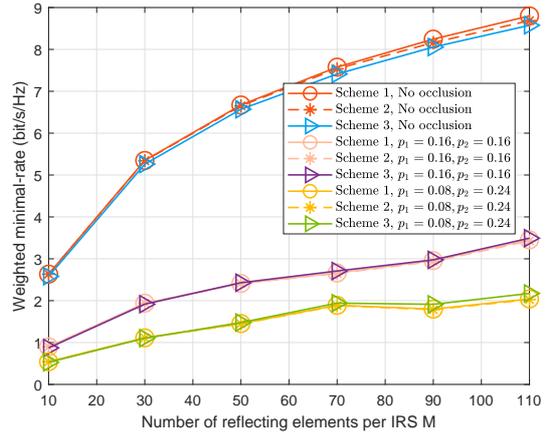}}
	\caption{Weighted minimal-rate against the number of reflecting elements per IRS under the condition that the channels may be blocked.
	}
	\label{zhedang2}
\end{figure}

According to Fig. \ref{ds1} and Fig. \ref{ds2}, the additional gain is approximately at its maximum when $M_1=M_2$. So we assume $M_1=M_2=M$. Fig. \ref{zhedang1}-\ref{zhedang2} show that if any channel (except for the direct path from BS to users) is not blocked or the probability of being blocked for the equal length is the same, the difference between the three schemes is almost negligible.
When the probabilities are different, that is, $p_1\neq p_2$, the impact of the inter-IRS channels becomes significant. The reason is that the pass-loss of the cascaded channel is in the form of product rather than addition, resulting in the channel quality of $\mathbf{g}_{2,k}^{H}\mathbf{\Phi }_{2}\mathbf{\Lambda}_{1,2}\mathbf{\Phi }_{1}\mathbf{H}_{1}$ several orders of magnitude worse than those of other channels.
%Therefore, in the former case, the optimization priority of $\mathbf{g}_{2,k}^{H}\mathbf{\Phi }_{2}\mathbf{\Lambda}_{1,2}\mathbf{\Phi }_{1}\mathbf{H}_{1}$ is relatively low.
As a result, the resource allocated to this cascaded channel is relatively low.

\subsection{Weighted Sum-Rate Maximization}\label{5A}

Starting from this subsection, it is assumed that no channel (except for the direct path from BS to users) is blocked and all IRSs belong to the same type, so no distinction is made between IRS1 and IRS2.

\begin{figure}[htbp]
	\centerline{\includegraphics[width=0.45\textwidth]{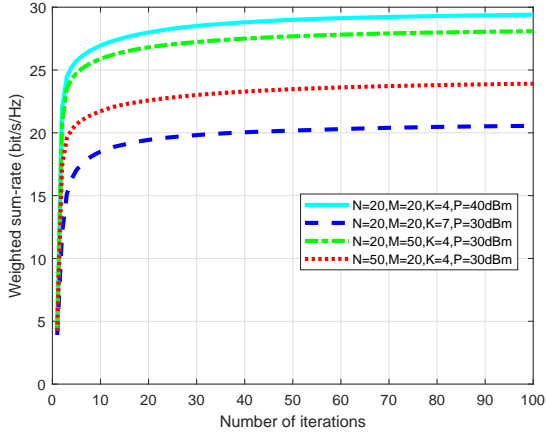}}
	\caption{Weighted sum-rate against the number of iterations.}
	\label{i}
\end{figure}

Fig. \ref{i} shows the convergence behavior of the DOMALO algorithm. It is observed that the weighted sum-rate increases monotonically over iterations, and the algorithm still guarantees convergence even when the system parameters change, which is consistent with \ref{Convergence1}. Taking into account the running speed and performance of the DOMALO algorithm, in this subsection, we set the maximum iterations $T_{max}$ to $ 30 $.

\begin{figure}[htbp]
	\centerline{\includegraphics[width=0.45\textwidth]{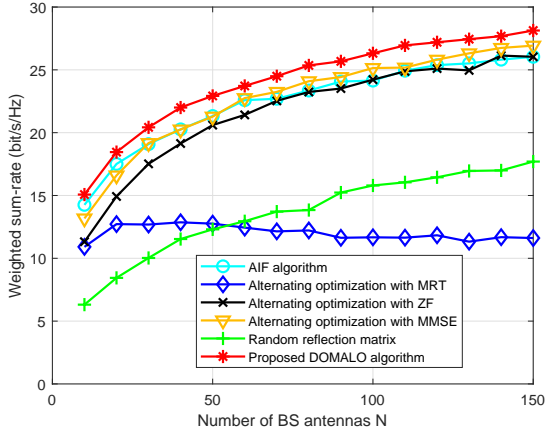}}
	\caption{Weighted sum-rate against the number of BS antennas.
	%\textcolor{red}{name the red line as "Proposed DAMO algorithm"}
	}
	\label{n}
\end{figure}

With $ M=20 $, $ K=4 $ and $ P=30\mathrm{dBm} $, the weighted sum-rate results against the number of BS antennas are shown in Fig. \ref{n}. As the number of BS antennas increases, the weighted sum-rate increases monotonically since more BS antennas can improve the channel conditions. However, the upward trend is getting slower and slower due to the fixed transmission power constraint, which is set to $ 30\mathrm{dBm} $. Moreover, the proposed algorithm has an advantage over other benchmark schemes for any number of BS antennas.

\begin{figure}[htbp]
	\centerline{\includegraphics[width=0.45\textwidth]{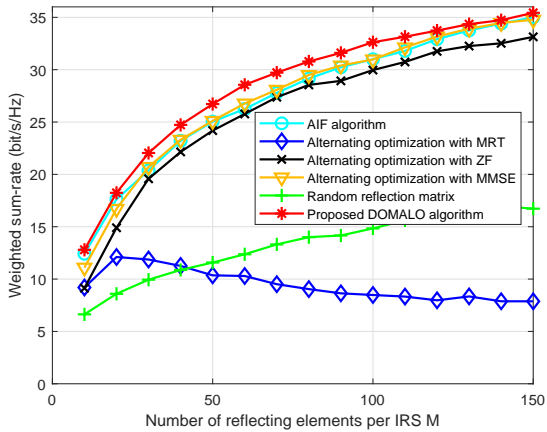}}
	\caption{Weighted sum-rate against the number of reflecting elements per IRS.}
	\label{m}
\end{figure}

In Fig. \ref{m}, the weighted sum-rate against the number of reflecting elements is presented, with  the number of BS antennas fixed at $ N=20 $, the number of users $ K=4 $, and the maximum transmission power is set as $ P=30\mathrm{dBm} $. Note that as the number of IRS elements grows, the sum-rate also increases, except for the alternating optimization scheme with MRT. Again the proposed DOMALO algorithm has provided the best performance in this setting.
\begin{figure}[htbp]
	\centerline{\includegraphics[width=0.45\textwidth]{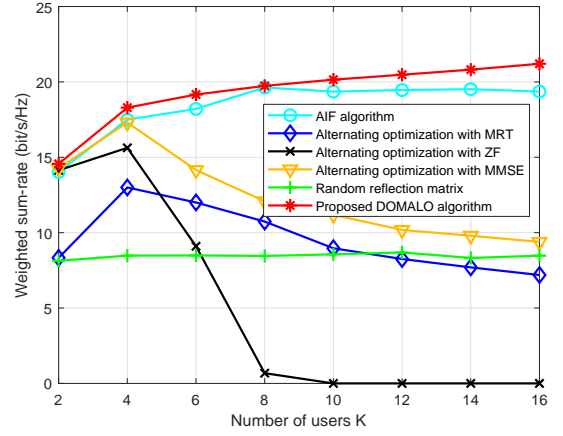}}
	\caption{Weighted sum-rate against the number of users K.}
	\label{k}
\end{figure}

Fig. \ref{k} shows the weighted sum-rate against the number of users with $ M=20 $, $ N=20 $ and $ P=30\mathrm{dBm}$. It can be seen that the proposed DOMALO algorithm has achieved the best performance among all the considered schemes. Moreover, notice that the weighted sum-rates of MRT based alternating optimization, MMSE based alternating optimization, and ZF based alternating optimization are reduced when the number of users increases and the weighted sum-rate of ZF based alternating optimization declines much faster than MRT and MMSE. It should be noted that the result of ZF based alternating optimization is invalid for $ K>8 $, as explained below.

For $  \mathbf{H}_{ZF}\mathbf{H}_{ZF}^H\in\mathbb{C}^{K\times K} $, its rank can be calculated by $  \mathrm{rank}\left( \mathbf{H}_{ZF}\mathbf{H}_{ZF}^H\right) =\min\left\lbrace N, SM, K, L_{BI}\right\rbrace $, where $ L_{BI} $ denotes the sum of the number of paths from BS to all IRSs. Specifically, let \{$ N=20, S=2, M=20, L_{BI}=S(L+1)=8 $\}, and then if $ K>8 $, $ \mathrm{rank}\left( \mathbf{H}_{ZF}\mathbf{H}_{ZF}^H\right)=8 $, and $  \mathbf{H}_{ZF}\mathbf{H}_{ZF}^H $ becomes singular, and therefore  (\ref{ZF}) is invalid.

\begin{figure}[htbp]
	\centerline{\includegraphics[width=0.45\textwidth]{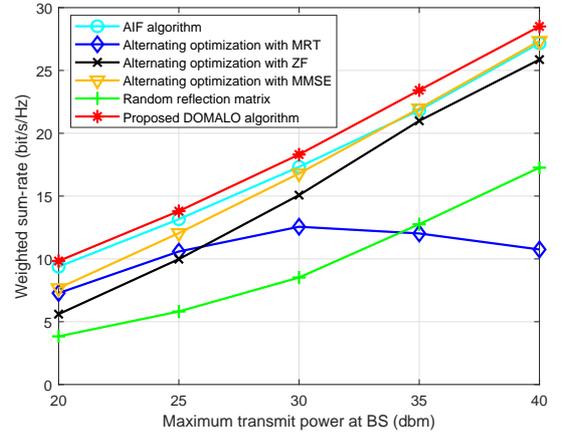}}
	\caption{Weighted sum-rate against the maximum transmission power at BS.}
	\label{d}
\end{figure}

With $ N=20 $, $ M=20$ and $ K=4$, Fig. \ref{d} shows that, with transmission power increasing from $20 \mathrm{dBm}$ to $40 \mathrm{dBm}$, all of the sum-rates except MRT exhibit a similar  upward trend, while the proposed DOMALO algorithm has the best performance.

From Fig. \ref{n} to Fig. \ref{d}, we can find that the alternating optimization with MRT performs worse with the increase of the number of BS antennas, reflecting elements, users, and maximum transmission power at BS. On the one hand, using MRT theory to update $ \mathbf{V} $ is self-centered because it does not consider user interference. This will cause the algorithm performance to be much worse than other benchmark methods when updating $ \mathbf{V} $, which in turn cause the algorithm to require more iterations for convergence. However, $ T_{max}  $ has been set to a specific value so that the MRT-based benchmark's output solution may not be a converged one. On the other hand, according to (\ref{MRT}), the increase of $ P $ makes the step of updating $ \mathbf{V} $ more influential, making this benchmark less-stable. The growth of $ N $, $ M $, and $ K $ makes the energy in the interference part much larger after initializing $ \mathbf{V} $ and $ \mathbf{\Phi} $, and then convergence requires more iterations than before. Hence, the output solution of the MRT-based benchmark may even lead to worse performance.

\begin{figure}[htbp]
	\centerline{\includegraphics[width=0.45\textwidth]{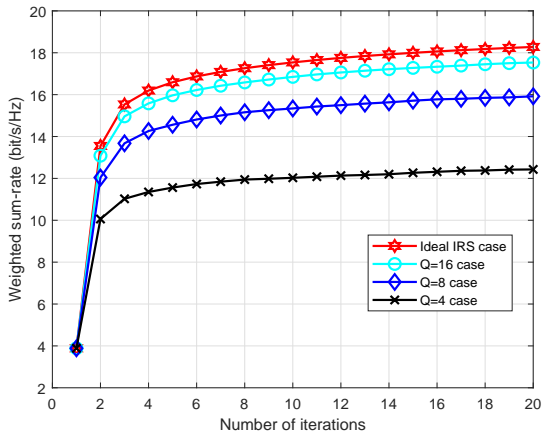}}
	\caption{Weighted sum-rate against the number of quantization levels. %\textcolor{red}{replace $L$ using $Q$}
	}
	\label{f}
\end{figure}

At last, we consider the non-ideal IRS case. With $ N=20, M=20 $, $ K=4 $ and $P=30\mathrm{dBm}$, the performance against the quantization level is shown in Fig. \ref{f}. As the number of quantization levels $ Q $ increases, the weight sum-rate increases. Moreover, the convergence of  \textbf{Algorithm 2} is still guaranteed. When $ Q $ approaches infinity, the performance will coincide with the ideal performance.

\subsection{Weighted Minimal-Rate Maximization}

\begin{figure}[htbp]
	\centerline{\includegraphics[width=0.45\textwidth]{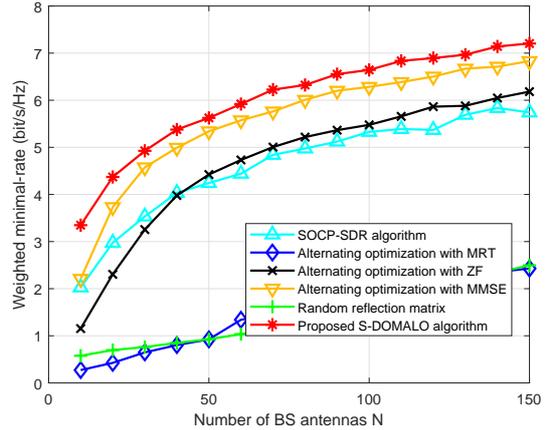}}
	\caption{Weighted minimal-rate against the number of BS antennas.}
	\label{n2}
\end{figure}

\begin{figure}[htbp]
	\centerline{\includegraphics[width=0.45\textwidth]{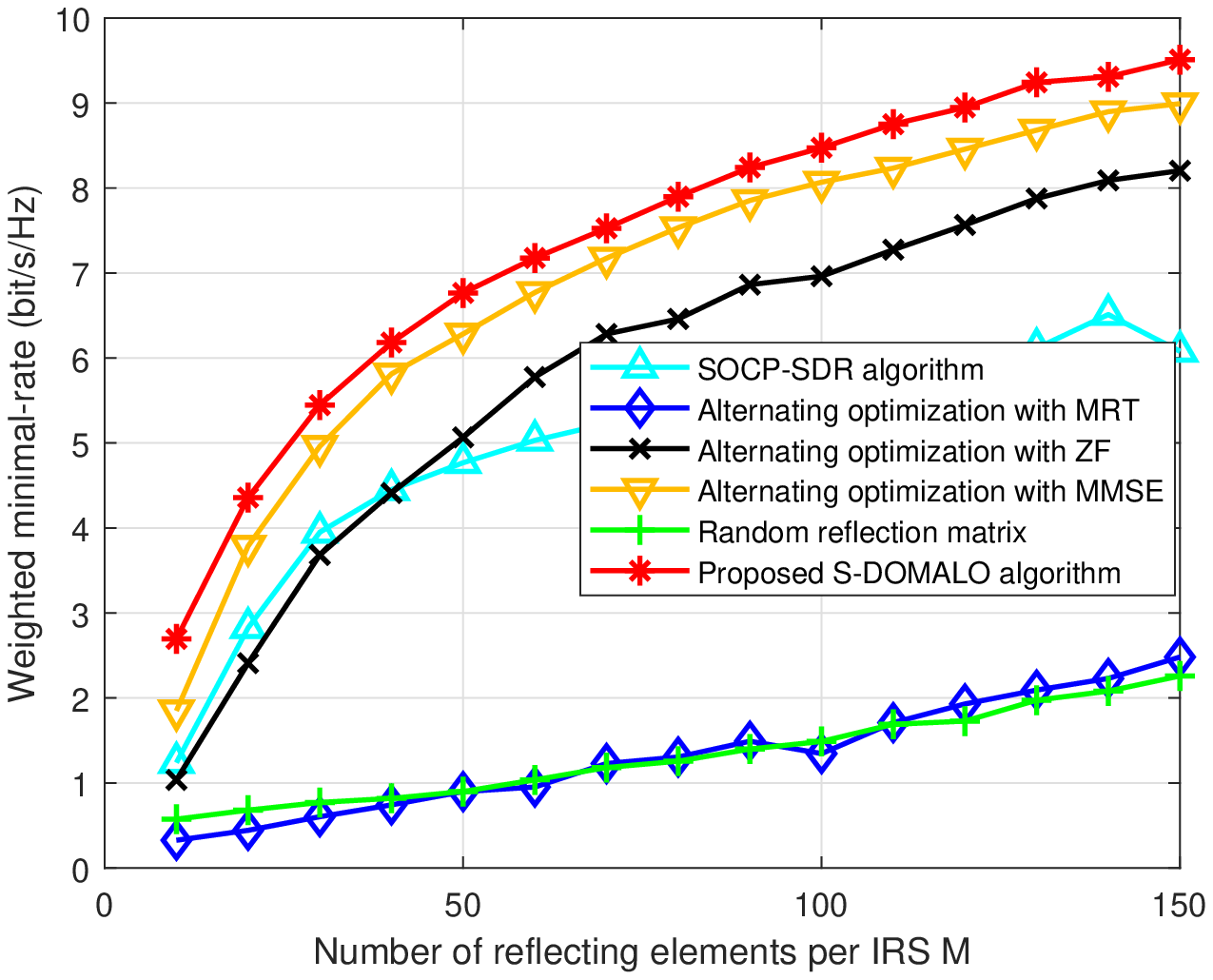}}
	\caption{Weighted minimal-rate against the number of reflecting elements per IRS.}
	\label{m2}
\end{figure}

The simulation setup is the same as in Section  \ref{5A}.
Figs. \ref{n2}-\ref{m2} show that the proposed S-DOMALO algorithm has considerably outperformed the other five benchmark schemes for all cases, where $ N $ and $ M $ take values from $10$ to $150$. In addition to the S-DOMALO algorithm, MMSE based alternating optimization can also provide  a relatively good performance. It is also observed that the performance of the six algorithms gets better gradually as $ N $ or $ M $ becomes larger. We can see from Fig.  \ref{n2} that alternating optimization with ZF and the SOCP-SDR algorithm achieve similar performances. In contrast, the performance of alternating optimization with ZF in Fig. \ref{m2} is better than that of the SOCP-SDR algorithm. The reason is the performance loss caused by the SDR method becomes larger when $ M $ increases, even if we generate 100,000 vectors in the Gaussian randomization procedure to reduce the performance loss while ensuring rank one constraint.


\begin{figure}[htbp]
	\centerline{\includegraphics[width=0.45\textwidth]{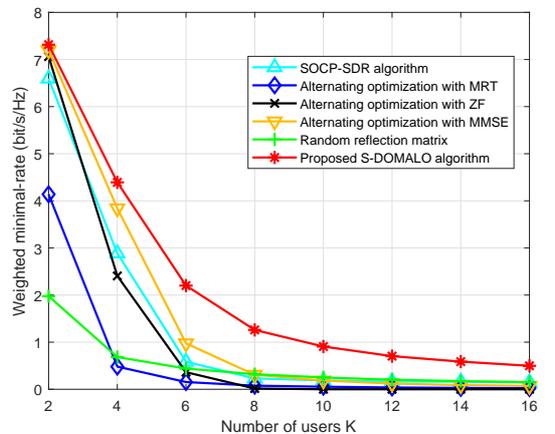}}
	\caption{Weighted minimal-rate against the number of users K.}
	\label{k2}
\end{figure}
\begin{figure}[htbp]
	\centerline{\includegraphics[width=0.45\textwidth]{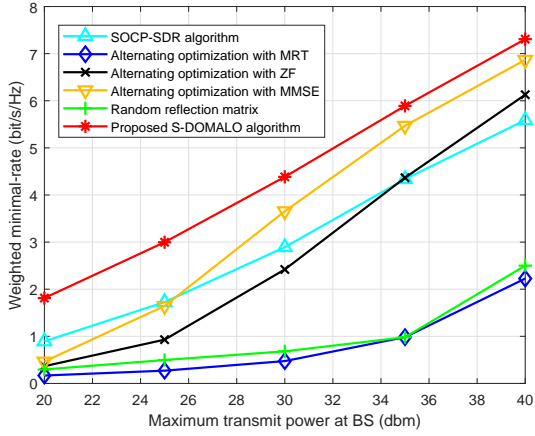}}
	\caption{Weighted minimal-rate against the maximum transmission power at BS.}
	\label{d2}
\end{figure}
\begin{figure}[htbp]
	\centerline{\includegraphics[width=0.45\textwidth]{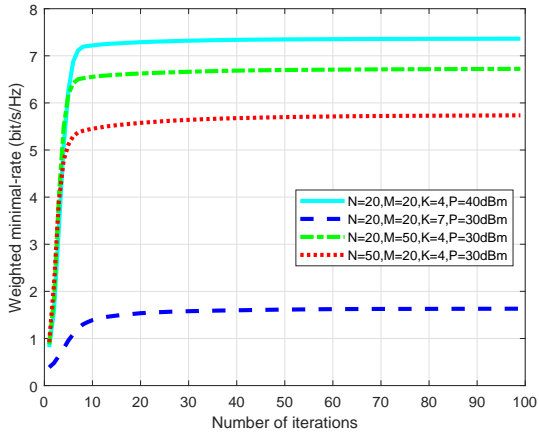}}
	\caption{Weighted minimal-rate against the number of iterations.}
	\label{i2}
\end{figure}

Fig. \ref{k2} shows the weighted minimal-rate against the number of users where $ M=20 $, $ N=20 $ and $ P=30\mathrm{dBm} $. It can be seen that the proposed S-DOMALO algorithm has achieved the best performance again. In addition, the overall trend of the curves is consistent with the expectation as max-min fairness needs to consider the worst channel state among all users. This also causes the trend of the curve of the S-DOMALO algorithm to be opposite to that of the DOMALO algorithm. Fig. \ref{d2} shows the weighted minimal-rate against the maximum transmission power. Same as in Fig. \ref{d}, the proposed S-DOMALO algorithm exhibits the best performance among all values of $P$. Specifically, the convergence of the S-DOMALO algorithm is shown by plotting the weighted minimal-rate against the number of iterations in Fig. \ref{i2}. It can be observed that it has converged quickly  within $ 20 $ iterations.

\begin{figure}[htbp]
	\centerline{\includegraphics[width=0.45\textwidth]{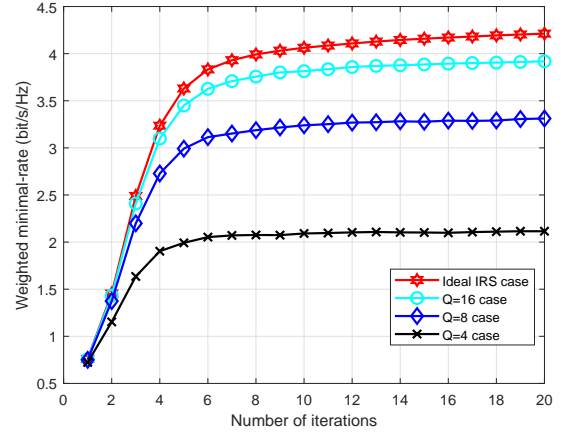}}
	\caption{Weighted minimal-rate against quantization levels.}
	\label{l2}
\end{figure}

Finally, the weighted minimal-rate against quantization levels is shown in Fig. \ref{l2}. Similar to Fig. \ref{f}, increasing quantization levels will improve the weighted minimal-rate, and the upper bound is the red line, i.e., the ideal case. %Considering the hardware implementation of IRS \cite{dai2020reconfigurable}, research on non-ideal IRS is of more practical significance.
Because the convergence is maintained, \textbf{Algorithm 3} is still effective under this condition.

\section{Conclusion}\label{Conclusion}
In this paper, two weighted rate maximization problems are formulated for a multi-user multi-IRS system. Both problems are solved by jointly optimizing the BS beamforming matrix and the IRS reflection matrix. The DOMALO algorithm is first proposed to solve the weighted sum-rate maximization problem based on the manifold optimization approach. To ensure fairness in resource allocation among different users, the S-DOMALO algorithm is developed to solve the max-min problem.
The multi IRS inter-IRS channel case is also considered, which provides additional cooperative gain.
The impact of the inter-IRS channels becomes significant only when the probability of a particular channel being blocked is relatively large.
Moreover, the quantization effect on the phase response of IRS is taken into consideration for the non-ideal IRS case. Simulation results demonstrated that the proposed manifold optimization algorithms have outperformed several benchmark schemes.

\bibliographystyle{IEEEtran}
\bibliography{ref}

\begin{thebibliography}{10}
\providecommand{\url}[1]{#1}
\csname url@samestyle\endcsname
\providecommand{\newblock}{\relax}
\providecommand{\bibinfo}[2]{#2}
\providecommand{\BIBentrySTDinterwordspacing}{\spaceskip=0pt\relax}
\providecommand{\BIBentryALTinterwordstretchfactor}{4}
\providecommand{\BIBentryALTinterwordspacing}{\spaceskip=\fontdimen2\font plus
\BIBentryALTinterwordstretchfactor\fontdimen3\font minus
  \fontdimen4\font\relax}
\providecommand{\BIBforeignlanguage}[2]{{%
\expandafter\ifx\csname l@#1\endcsname\relax
\typeout{** WARNING: IEEEtran.bst: No hyphenation pattern has been}%
\typeout{** loaded for the language `#1'. Using the pattern for}%
\typeout{** the default language instead.}%
\else
\language=\csname l@#1\endcsname
\fi
#2}}
\providecommand{\BIBdecl}{\relax}
\BIBdecl

\bibitem{9580434}
L.~Zhang, Q.~Wang, and H.~Wang, ``Multiple intelligent reflecting surface aided
  multi-user weighted sum-rate maximization using manifold optimization,'' in
  \emph{2021 IEEE/CIC International Conference on Communications in China
  (ICCC)}, 2021, pp. 364--369.

\bibitem{towards}
Q.~Wu and R.~Zhang, ``Towards smart and reconfigurable environment: Intelligent
  reflecting surface aided wireless network,'' \emph{IEEE Communications
  Magazine}, vol.~58, no.~1, pp. 106--112, Nov. 2019.

\bibitem{9509294}
D.~C. Nguyen, M.~Ding, P.~N. Pathirana, A.~Seneviratne, J.~Li, D.~Niyato,
  O.~Dobre, and H.~V. Poor, ``{6G} internet of things: A comprehensive
  survey,'' \emph{IEEE Internet of Things Journal}, vol.~9, no.~1, pp.
  359--383, 2022.

\bibitem{MIMO1}
E.~Bj{\"o}rnson and L.~Sanguinetti, ``Power scaling laws and near-field
  behaviors of massive {MIMO} and intelligent reflecting surfaces,'' \emph{IEEE
  Open Journal of the Communications Society}, vol.~1, pp. 1306--1324, Sep.
  2020.

\bibitem{li2019joint}
Y.~Li, M.~Jiang, Q.~Zhang, and J.~Qin, ``Joint beamforming design in
  multi-cluster {MISO} {NOMA} intelligent reflecting surface-aided downlink
  communication networks,'' \emph{arXiv preprint arXiv:1909.06972}, Sep. 2019.

\bibitem{wu2020joint}
Q.~Wu and R.~Zhang, ``Joint active and passive beamforming optimization for
  intelligent reflecting surface assisted {SWIPT} under {QoS} constraints,''
  \emph{IEEE Journal on Selected Areas in Communications}, Jul. 2020.

\bibitem{cao2020intelligent}
Y.~Cao, T.~Lv, and W.~Ni, ``Intelligent reflecting surface aided multi-user
  mm{W}ave communications for coverage enhancement,'' in \emph{2020 IEEE 31st
  Annual International Symposium on Personal, Indoor and Mobile Radio
  Communications}.\hskip 1em plus 0.5em minus 0.4em\relax IEEE, Oct. 2020, pp.
  1--6.

\bibitem{peng2020multiuser}
Z.~Peng, Z.~Zhang, C.~Pan, L.~Li, and A.~L. Swindlehurst, ``Multiuser
  full-duplex two-way communications via intelligent reflecting surface,''
  \emph{IEEE Transactions on Signal Processing}, vol.~69, pp. 837--851, Jan.
  2021.

\bibitem{cui2019secure}
M.~Cui, G.~Zhang, and R.~Zhang, ``Secure wireless communication via intelligent
  reflecting surface,'' \emph{IEEE Wireless Communications Letters}, vol.~8,
  no.~5, pp. 1410--1414, May 2019.

\bibitem{dong2020secure}
L.~Dong and H.-M. Wang, ``Secure {MIMO} transmission via intelligent reflecting
  surface,'' \emph{IEEE Wireless Communications Letters}, vol.~9, no.~6, pp.
  787--790, Jan. 2020.

\bibitem{9612585}
B.~Li, W.~Wu, Y.~Li, and W.~Zhao, ``Intelligent reflecting surface and
  artificial noise assisted secure transmission of {MEC} system,'' \emph{IEEE
  Internet of Things Journal}, pp. 1--1, 2021.

\bibitem{9416177}
X.~Wang, Z.~Fei, Z.~Zheng, and J.~Guo, ``Joint waveform design and passive
  beamforming for {RIS}-assisted dual-functional radar-communication system,''
  \emph{IEEE Transactions on Vehicular Technology}, vol.~70, no.~5, pp.
  5131--5136, Apr. 2021.

\bibitem{9264225}
X.~Wang, Z.~Fei, J.~Guo, Z.~Zheng, and B.~Li, ``{RIS}-assisted spectrum sharing
  between {MIMO} radar and {MU-MISO} communication systems,'' \emph{IEEE
  Wireless Communications Letters}, vol.~10, no.~3, pp. 594--598, Nov. 2021.

\bibitem{9305278}
S.~Gong, Z.~Yang, C.~Xing, J.~An, and L.~Hanzo, ``Beamforming optimization for
  intelligent reflecting surface-aided {SWIPT} {IoT} networks relying on
  discrete phase shifts,'' \emph{IEEE Internet of Things Journal}, vol.~8,
  no.~10, pp. 8585--8602, 2021.

\bibitem{feng2020deep}
K.~Feng, Q.~Wang, X.~Li, and C.-K. Wen, ``Deep reinforcement learning based
  intelligent reflecting surface optimization for {MISO} communication
  systems,'' \emph{IEEE Wireless Communications Letters}, vol.~9, no.~5, pp.
  745--749, Jan. 2020.

\bibitem{8746155}
Y.~Han, W.~Tang, S.~Jin, C.-K. Wen, and X.~Ma, ``Large intelligent
  surface-assisted wireless communication exploiting statistical {CSI},''
  \emph{IEEE Transactions on Vehicular Technology}, vol.~68, no.~8, pp.
  8238--8242, Jun. 2019.

\bibitem{9013288}
H.~Guo, Y.-C. Liang, J.~Chen, and E.~G. Larsson, ``Weighted sum-rate
  maximization for intelligent reflecting surface enhanced wireless networks,''
  in \emph{2019 IEEE Global Communications Conference (GLOBECOM)}, Feb. 2019,
  pp. 1--6.

\bibitem{chen2019sum}
W.~Chen, X.~Ma, Z.~Li, and N.~Kuang, ``Sum-rate maximization for intelligent
  reflecting surface based {Terahertz} communication systems,'' in \emph{2019
  IEEE/CIC International Conference on Communications Workshops in China (ICCC
  Workshops)}.\hskip 1em plus 0.5em minus 0.4em\relax IEEE, Aug. 2019, pp.
  153--157.

\bibitem{9531458}
J.-C. Chen, ``Machine learning-inspired algorithmic framework for intelligent
  reflecting surface-assisted wireless systems,'' \emph{IEEE Transactions on
  Vehicular Technology}, pp. 1--1, Sep. 2021.

\bibitem{tang2020joint}
Y.~Tang, G.~Ma, H.~Xie, J.~Xu, and X.~Han, ``Joint transmit and reflective
  beamforming design for {IRS}-assisted multiuser {MISO SWIPT} systems,'' in
  \emph{ICC 2020-2020 IEEE International Conference on Communications
  (ICC)}.\hskip 1em plus 0.5em minus 0.4em\relax IEEE, Jul. 2020, pp. 1--6.

\bibitem{kammoun2020asymptotic}
A.~Kammoun, A.~Chaaban, M.~Debbah, M.-S. Alouini \emph{et~al.}, ``Asymptotic
  max-min {SINR} analysis of reconfigurable intelligent surface assisted {MISO}
  systems,'' \emph{IEEE Transactions on Wireless Communications}, vol.~19,
  no.~12, pp. 7748--7764, Apr. 2020.

\bibitem{xie2020max}
H.~Xie, J.~Xu, and Y.-F. Liu, ``Max-min fairness in {IRS}-aided multi-cell
  {MISO} systems with joint transmit and reflective beamforming,'' \emph{IEEE
  Transactions on Wireless Communications}, Oct. 2020.

\bibitem{zhao2020two}
M.-M. Zhao, Q.~Wu, M.-J. Zhao, and R.~Zhang, ``Two-timescale beamforming
  optimization for intelligent reflecting surface enhanced wireless network,''
  in \emph{2020 IEEE 11th Sensor Array and Multichannel Signal Processing
  Workshop (SAM)}.\hskip 1em plus 0.5em minus 0.4em\relax IEEE, Jun. 2020, pp.
  1--5.

\bibitem{yang2020irs}
Y.~Yang, S.~Zhang, and R.~Zhang, ``{IRS}-enhanced {OFDMA}: Joint resource
  allocation and passive beamforming optimization,'' \emph{IEEE Wireless
  Communications Letters}, vol.~9, no.~6, pp. 760--764, Jan. 2020.

\bibitem{9497353}
M.~He, W.~Xu, H.~Shen, G.~Xie, C.~Zhao, and M.~Di~Renzo, ``Cooperative
  {M}ulti-{RIS} communications for wideband mmwave {MISO-OFDM} systems,''
  \emph{IEEE Wireless Communications Letters}, vol.~10, no.~11, pp. 2360--2364,
  2021.

\bibitem{yang2020intelligent}
G.~Yang, X.~Xu, and Y.-C. Liang, ``Intelligent reflecting surface assisted
  non-orthogonal multiple access,'' in \emph{2020 IEEE Wireless Communications
  and Networking Conference (WCNC)}.\hskip 1em plus 0.5em minus 0.4em\relax
  IEEE, Jun. 2020, pp. 1--6.

\bibitem{9362274}
B.~Zheng, C.~You, and R.~Zhang, ``Double-{IRS} assisted multi-user {MIMO}:
  {Cooperative} passive beamforming design,'' \emph{IEEE Transactions on
  Wireless Communications}, vol.~20, no.~7, pp. 4513--4526, 2021.

\bibitem{9417491}
Y.~Xiu, W.~Sun, J.~Wu, G.~Gui, N.~Wei, and Z.~Zhang, ``Sum-rate maximization in
  distributed intelligent reflecting surfaces-aided mmwave communications,'' in
  \emph{2021 IEEE Wireless Communications and Networking Conference (WCNC)},
  2021, pp. 1--6.

\bibitem{li2020manifold}
J.~Li, G.~Liao, Y.~Huang, and A.~Nehorai, ``Manifold optimization for joint
  design of {MIMO-STAP} radars,'' \emph{IEEE Signal Processing Letters}, Oct.
  2020.

\bibitem{9322519}
T.~Lin, X.~Yu, Y.~Zhu, and R.~Schober, ``Channel estimation for intelligent
  reflecting surface-assisted millimeter wave {MIMO} systems,'' in
  \emph{GLOBECOM 2020 - 2020 IEEE Global Communications Conference}, Jan. 2020,
  pp. 1--6.

\bibitem{hong2020semi}
X.~Hong, J.~Gao, and S.~Chen, ``Semi-blind joint channel estimation and data
  detection on sphere manifold for {MIMO} with high-order {QAM} signaling,''
  \emph{Journal of the Franklin Institute}, Jun. 2020.

\bibitem{8855810}
X.~Yu, D.~Xu, and R.~Schober, ``{MISO} wireless communication systems via
  intelligent reflecting surfaces : {(Invited Paper)},'' in \emph{2019 IEEE/CIC
  International Conference on Communications in China (ICCC)}, 2019, pp.
  735--740.

\bibitem{douik2019manifold}
A.~Douik and B.~Hassibi, ``Manifold optimization over the set of doubly
  stochastic matrices: A second-order geometry,'' \emph{IEEE Transactions on
  Signal Processing}, vol.~67, no.~22, pp. 5761--5774, Oct. 2019.

\bibitem{bjornson2021reconfigurable}
E.~Bj{\"o}rnson, H.~Wymeersch, B.~Matthiesen, P.~Popovski, L.~Sanguinetti, and
  E.~de~Carvalho, ``Reconfigurable intelligent surfaces: A signal processing
  perspective with wireless applications,'' \emph{arXiv preprint
  arXiv:2102.00742}, Feb. 2021.

\bibitem{cvx}
M.~Grant and S.~Boyd, ``{CVX}: Matlab software for disciplined convex
  programming, version 2.1,'' \url{http://cvxr.com/cvx}, Mar. 2014.

\bibitem{shen2018fractional}
K.~Shen and W.~Yu, ``Fractional programming for communication systems—part
  {II}: Uplink scheduling via matching,'' \emph{IEEE Transactions on Signal
  Processing}, vol.~66, no.~10, pp. 2631--2644, Mar. 2018.

\bibitem{9459505}
Z.~Zhang and L.~Dai, ``A joint precoding framework for wideband reconfigurable
  intelligent surface-aided cell-free network,'' \emph{IEEE Transactions on
  Signal Processing}, vol.~69, pp. 4085--4101, Jun. 2021.

\bibitem{al}
P.-A. Absil, R.~Mahony, and R.~Sepulchre, \emph{Optimization algorithms on
  matrix manifolds}.\hskip 1em plus 0.5em minus 0.4em\relax Princeton
  University Press, Apr. 2009.

\bibitem{boumal2020introduction}
N.~Boumal, ``An introduction to optimization on smooth manifolds,''
  \emph{Available online, May}, 2020.

\bibitem{hu2020brief}
J.~Hu, X.~Liu, Z.-W. Wen, and Y.-X. Yuan, ``A brief introduction to manifold
  optimization,'' \emph{Journal of the Operations Research Society of China},
  vol.~8, no.~2, pp. 199--248, Jun. 2020.

\bibitem{barros1996new}
A.~Barros, J.~Frenk, S.~Schaible, and S.~Zhang, ``A new algorithm for
  generalized fractional programs,'' \emph{Mathematical Programming}, vol.~72,
  no.~2, pp. 147--175, Feb. 1996.

\bibitem{crouzeix1985algorithm}
J.~Crouzeix, J.~Ferland, and S.~Schaible, ``An algorithm for generalized
  fractional programs,'' \emph{Journal of Optimization Theory and
  Applications}, vol.~47, no.~1, pp. 35--49, Sep. 1985.

\bibitem{xu2001smoothing}
S.~Xu, ``Smoothing method for minimax problems,'' \emph{Computational
  Optimization and Applications}, vol.~20, no.~3, pp. 267--279, Dec. 2001.

\bibitem{polak2003algorithms}
E.~Polak, J.~Royset, and R.~Womersley, ``Algorithms with adaptive smoothing for
  finite minimax problems,'' \emph{Journal of Optimization Theory and
  Applications}, vol. 119, no.~3, pp. 459--484, Dec. 2003.

\bibitem{el2014spatially}
O.~El~Ayach, S.~Rajagopal, S.~Abu-Surra, Z.~Pi, and R.~W. Heath, ``Spatially
  sparse precoding in millimeter wave {MIMO} systems,'' \emph{IEEE transactions
  on wireless communications}, vol.~13, no.~3, pp. 1499--1513, Jan. 2014.

\bibitem{cho2010mimo}
Y.~S. Cho, J.~Kim, W.~Y. Yang, and C.~G. Kang, \emph{{MIMO-OFDM} wireless
  communications with MATLAB}.\hskip 1em plus 0.5em minus 0.4em\relax John
  Wiley \& Sons, Aug. 2010.

\bibitem{9348016}
C.~Kai, W.~Ding, and W.~Huang, ``Max-{M}in fairness in {IRS}-aided {MISO}
  broadcast channel via joint transmit and reflective beamforming,'' in
  \emph{GLOBECOM 2020 - 2020 IEEE Global Communications Conference}, Dec. 2020,
  pp. 1--6.

\end{thebibliography}

\end{document}